\documentclass[12pt]{article}
\usepackage{amsmath}
\usepackage{bm}
\usepackage{amssymb}
\usepackage{graphicx}
\usepackage{natbib}
\usepackage{url} 
\usepackage{xcolor}

\newcommand{\colorcircle}[2][]{%
  \tikz[baseline=-0.5ex] \node [circle, draw, black, very thick, 
  fill=#2, inner sep=1mm, minimum size=0.3cm, #1] {};
}

\newcommand{\ddd}{0.9}
\newcommand{\dd}{1.2}
\newcommand{\xxx}{1.2}
\newcommand{\yyy}{1.2}

\usepackage{booktabs}
\usepackage{subcaption}
\usepackage{bm}
\usepackage{bbm}
\usepackage{tikz}
\usepackage{array}
\usepackage{multirow}
\newcommand\independent{\protect\mathpalette{\protect\independenT}{\perp}}
\def\independenT#1#2{\mathrel{\rlap{$#1#2$}\mkern2mu{#1#2}}}

\usepackage{longtable}
\usepackage{standalone}
\newcommand{\blind}{0}

\begin{document}

\def\spacingset#1{\renewcommand{\baselinestretch}%
{#1}\small\normalsize} \spacingset{1}


\if0\blind
{
  \title{\bf Bayesian Causal Effect Estimation for Categorical Data using Staged Tree Models}
  \author{
   Andrea Cremaschi
   \\
   \vspace{0.4cm}
    School of Science and Technology, IE University, Madrid, Spain\\
   Manuele Leonelli
   \\
   \vspace{0.4cm}
    School of Science and Technology, IE University, Madrid, Spain
    \\
    Gherardo Varando \\
    Department of Statistics and Operational Research,\\ University of Valencia, Valencia, Spain
    }
  \maketitle
} \fi

\if1\blind
{
  \bigskip
  \bigskip
  \bigskip
  \begin{center}
    {\LARGE\bf Title}
\end{center}
  \medskip
} \fi

\bigskip

\begin{abstract}
We propose a fully Bayesian approach for causal inference with multivariate categorical data based on staged tree models, a class of probabilistic graphical models capable of representing asymmetric and context-specific dependencies. To account for uncertainty in both structure and parameters, we introduce a flexible family of prior distributions over staged trees. These include product partition models to encourage parsimony, a novel distance-based prior to promote interpretable dependence patterns, and an extension that incorporates continuous covariates into the learning process. Posterior inference is achieved via a tailored Markov Chain Monte Carlo algorithm with split-and-merge moves, yielding posterior samples of staged trees from which average treatment effects and uncertainty measures are derived. Posterior summaries and uncertainty measures are obtained via techniques from the Bayesian nonparametrics literature. Two case studies on electronic fetal monitoring and cesarean delivery and on anthracycline therapy and cardiac dysfunction in breast cancer illustrate the methods.
\end{abstract}

{\it Keywords:}  Causal inference, Context-specific independence, Markov chain Monte Carlo, Product partition models, Staged trees

\newpage
\spacingset{1.2} 

\section{Introduction}
Understanding and quantifying causal effects is a central goal across many scientific disciplines, requiring the integration of statistical modeling, domain-specific assumptions, and empirical data~\citep{hernancausal,pearl2009causal, pearl2016causal}. A key objective in this context is the estimation of the average treatment effect (ATE), which measures the expected change in an outcome under different treatment or intervention conditions~\citep{hernancausal}. Randomized controlled trials are widely regarded as the gold standard for estimating causal effects, as randomization effectively eliminates confounding~\citep{hernancausal}. However, ethical concerns, financial limitations, and practical constraints often render them infeasible. Consequently, substantial research has focused on developing robust methods for causal inference from observational data~\citep{hernancausal, runge2023causal}. 

Probabilistic graphical models, particularly directed acyclic graphs (DAGs), are widely used to represent causal assumptions and derive causal estimates. Their structured framework facilitates the identification of causal relationships, assessment of identifiability, and construction of valid estimators~\citep{huang2006identifiability}. However, DAGs are limited in their ability to represent asymmetry and context-specific dependencies~\citep{boutilier1996context}, which are often essential for capturing the complexity of real-world systems. Recent advances have shown that incorporating context-specific independencies can substantially improve the precision and reliability of causal effect identification, especially in observational studies~\citep{chen2024constrained, mokhtarian2022causal, tikka2019identifying}. By modeling these nuanced dependencies, graphical models that go beyond standard DAGs can offer a more refined and accurate representation of causal mechanisms. 

Staged trees have emerged as a powerful class of probabilistic graphical models, providing a flexible framework for representing asymmetry and for encoding conditional independencies that hold only in specific contexts~\citep{smith2008conditional, collazo2018chain}. In recent years, multiple efficient algorithms for their construction and analysis have been developed~\citep{leonelli2024structural,varando2024staged}, with open-source implementations available~\citep{Carli2020}. These advancements have demonstrated the utility of staged trees in the realms of causal discovery and inference, particularly in observational settings where asymmetry and context-specific dependencies play a critical role~\citep{cowell2014causal,gorgen2018discovery,leonelli2023context,thwaites2013causal,thwaites2010causal,varando2025causal}.
Staged trees are ideal for modeling multiple categorical variables observed simultaneously, a topic that has witnessed a strong interest in the last few years \citep{fop2017variable, argiento2025model, malsiner2025without}. Despite the prevalence of categorical data across many scientific disciplines, existing causal methods often assume multivariate continuous distributions~\citep{vonk2023disentangling}, leaving the discrete setting comparatively underexplored.

In this work, we consider the common scenario in which the true data-generating causal model is unknown. Rather than select a single model and then estimate effects, which can induce post-selection bias~\citep{berk2013valid}, we adopt a fully Bayesian approach that jointly learns staged-tree structure and causal effects. Related joint Bayesian estimators have been developed for DAGs~\citep{castelletti2021equivalence,castelletti2024joint,castelletti2024bayesian}, but they do not capture the context-specific asymmetries that staged trees encode. To promote parsimony and interpretability, we introduce a flexible class of prior distributions over staged tree models, moving beyond the near-exclusive reliance on uniform priors~\citep{freeman2011bayesian}. Drawing on the close connection between staged trees and clustering methods~\citep{shenvi2024beyond}, we first consider product partition models (PPMs)~\citep{quintana2003bayesian}, which provide a natural mechanism to favor simpler and more interpretable structures. Building on the formulation of \citet{cremaschi2023change}, we then propose a novel prior that incorporates pairwise similarities between configurations of the variables, favoring models that cluster together contexts with similar structural roles. This promotes staged trees that reflect coherent and interpretable patterns of dependence, while remaining parsimonious. Finally, we extend this prior formulation to incorporate continuous information using the Product Partition Model with covariates (PPMx) framework~\citep{muller2011product}. Covariates guide clustering through the prior while remaining external to the graphical representation~\citep{jewson2024graphical}. This is the first integration of continuous information into staged trees, contributing to the broader goal of developing interpretable graphical models for mixed data types~\citep{cai2022causal, cui2019learning}.

Posterior inference under the proposed framework is achieved via a tailored Markov Chain Monte Carlo (MCMC) algorithm that combines the approach by \citet{neal2000markov} for Dirichlet process mixtures with split-and-merge moves, enabling efficient exploration of the space of staged trees. The algorithm yields posterior samples of staged tree models from which ATEs and uncertainty measures can be derived using standard Bayesian tools. We also address the important challenge of summarizing the posterior distribution by selecting a single representative staged tree that concisely captures the dependence structure among the variables. To this end, we adapt Bayesian clustering techniques~\citep{wade2018bayesian} to summarize the posterior sample of partitions, and we further provide a visualization of model uncertainty using credible balls around the selected staged tree, offering insight into the stability of its inferred structure. 

Our methodology relates to recent Bayesian models for categorical data that capture heterogeneity~\citep{argiento2025model}, such as clustering of categorical distributions and nonparametric models for heterogeneous undirected graphs~\citep{barile2024bayesian}. These approaches, however, do not exploit staged trees’ explicit representation of context-specific and asymmetric dependencies. Another line of research related to the one discussed in this work concerns the integration of expert knowledge into Bayesian causal discovery. Since \citet{heckerman1995learning}, DAGs have provided a natural way to encode prior beliefs via structural scores and constraints~\citep{borboudakis2013scoring,castelo2000learning}, with extensions to partial or uncertain knowledge and to combining multiple information sources~\citep{amirkhani2016exploiting,werhli2007reconstructing}. We bring this perspective to staged trees by specifying informative priors that favor parsimonious, context-specific structures.

The paper is organized as follows. Section~\ref{sec:setup} introduces staged trees and causal inference. Section~\ref{sec:model} presents our modeling framework and prior specifications. Section~\ref{sec:posterio_inference} outlines the MCMC estimation algorithm and posterior summarization techniques. Section~\ref{sec:rhc} illustrates the practical utility of our method through two real-world case studies. Section~\ref{sec:concl} concludes with a discussion and future directions. A Supplementary Material file is available for this manuscript, with the details of the MCMC algorithm, proofs and additional figures and tables mentioned throughout the paper. Code and replication materials are available at \url{https://github.com/manueleleonelli/bayesian_stagedtrees}.

\section{The Setup}\label{sec:setup}

Let $[p] = \{0,\dots,p\}$ and $\bm{X} = (X_0, \dots, X_p) = (X_j)_{j \in [p]}$ be a sequence of categorical random variables with joint probability mass function $P$ and sample space $\mathbb{X} = \times_{j \in [p]} \mathbb{X}_j$, where each $\mathbb{X}_j$ is the finite set of possible values of $X_j$ and $\mathbb{X}$ indicates the resulting product space. For any subset $A \subseteq [p]$, we denote by $\bm{X}_A = (X_j)_{j \in A}$ the subvector of variables and by $\bm{x}_A = (x_j)_{j \in A} \in \mathbb{X}_A = \times_{j \in A} \mathbb{X}_j$ a generic configuration. For instance, $\bm{x}_{[i-1]} = (x_j)_{j \in [i-1]} \in \mathbb{X}_{[i-1]} = \times_{j \in [i-1]} \mathbb{X}_j$ denotes a generic configuration of the first $i$ variables, for any $i \in [p]$ with $i \neq 0$. We also write $\bm{X}_{-A} = \bm{X}_{[p] \setminus A}$.

\begin{figure}[t]
    \centering

    \begin{subfigure}{0.49\textwidth}
        \centering
        \resizebox{0.9\linewidth}{!}{%
\begin{tikzpicture}[auto, scale=14,
	X1_NA/.style={circle,inner sep=1mm,minimum size=0.4cm,draw,black,very thick,fill=white,text=black},
	X2_1/.style={circle,inner sep=1mm,minimum size=0.4cm,draw,black,very thick,fill=white,text=black},
	X3_1/.style={circle,inner sep=1mm,minimum size=0.4cm,draw,black,very thick,fill=white,text=black},
	X4_1/.style={circle,inner sep=1mm,minimum size=0.4cm,draw,black,very thick,fill=white,text=black},
	 leaf/.style={circle,inner sep=1mm,minimum size=0.2cm,draw,very thick,black,fill=gray,text=black}]

	\node [X1_NA] (root) at (0.000000, 0.500000)	{};
	\node [X2_1] (root-0) at (0.250000, 0.233333)	{};
	\node [X2_1] (root-1) at (0.250000, 0.766667)	{};
	\node [X3_1] (root-0-0) at (0.500000, 0.100000)	{};
	\node [X3_1] (root-0-1) at (0.500000, 0.366667)	{};
	\node [X3_1] (root-1-0) at (0.500000, 0.633333)	{};
	\node [X3_1] (root-1-1) at (0.500000, 0.900000)	{};
	\node [X4_1] (root-0-0-0) at (0.750000, 0.033333)	{};
	\node [X4_1] (root-0-0-1) at (0.750000, 0.166667)	{};
	\node [X4_1] (root-0-1-0) at (0.750000, 0.300000)	{};
	\node [X4_1] (root-0-1-1) at (0.750000, 0.433333)	{};
	\node [X4_1] (root-1-0-0) at (0.750000, 0.566667)	{};
	\node [X4_1] (root-1-0-1) at (0.750000, 0.700000)	{};
	\node [X4_1] (root-1-1-0) at (0.750000, 0.833333)	{};
	\node [X4_1] (root-1-1-1) at (0.750000, 0.966667)	{};
	\node [leaf] (root-0-0-0-0) at (1.000000, 0.000000)	{};
	\node [leaf] (root-0-0-0-1) at (1.000000, 0.066667)	{};
	\node [leaf] (root-0-0-1-0) at (1.000000, 0.133333)	{};
	\node [leaf] (root-0-0-1-1) at (1.000000, 0.200000)	{};
	\node [leaf] (root-0-1-0-0) at (1.000000, 0.266667)	{};
	\node [leaf] (root-0-1-0-1) at (1.000000, 0.333333)	{};
	\node [leaf] (root-0-1-1-0) at (1.000000, 0.400000)	{};
	\node [leaf] (root-0-1-1-1) at (1.000000, 0.466667)	{};
	\node [leaf] (root-1-0-0-0) at (1.000000, 0.533333)	{};
	\node [leaf] (root-1-0-0-1) at (1.000000, 0.600000)	{};
	\node [leaf] (root-1-0-1-0) at (1.000000, 0.666667)	{};
	\node [leaf] (root-1-0-1-1) at (1.000000, 0.733333)	{};
	\node [leaf] (root-1-1-0-0) at (1.000000, 0.800000)	{};
	\node [leaf] (root-1-1-0-1) at (1.000000, 0.866667)	{};
	\node [leaf] (root-1-1-1-0) at (1.000000, 0.933333)	{};
	\node [leaf] (root-1-1-1-1) at (1.000000, 1.000000)	{};

	\draw[->] (root) -- node [sloped,below]{$N^0$} (root-0);
	\draw[->] (root) -- node [sloped]{$N^1$} (root-1);
	\draw[->] (root-0) -- node [sloped,below]{$N_{(0)}^0$} (root-0-0);
	\draw[->] (root-0) -- node [sloped]{$N_{(0)}^1$} (root-0-1);
	\draw[->] (root-1) -- node [sloped,below]{$N_{(1)}^0$} (root-1-0);
	\draw[->] (root-1) -- node [sloped]{$N_{(1)}^1$} (root-1-1);
	\draw[->] (root-0-0) -- node [sloped,below]{$N_{(0,0)}^0$} (root-0-0-0);
	\draw[->] (root-0-0) -- node [sloped]{$N_{(0,0)}^1$} (root-0-0-1);
	\draw[->] (root-0-1) -- node [sloped,below]{$N_{(0,1)}^0$} (root-0-1-0);
	\draw[->] (root-0-1) -- node [sloped]{$N_{(0,1)}^1$} (root-0-1-1);
	\draw[->] (root-1-0) -- node [sloped,below]{$N_{(1,0)}^0$} (root-1-0-0);
	\draw[->] (root-1-0) -- node [sloped]{$N_{(1,0)}^1$} (root-1-0-1);
	\draw[->] (root-1-1) -- node [sloped,below]{$N_{(1,1)}^0$} (root-1-1-0);
	\draw[->] (root-1-1) -- node [sloped]{$N_{(1,1)}^1$} (root-1-1-1);
	\draw[->] (root-0-0-0) -- node [sloped,below]{$N_{(0,0,0)}^0$} (root-0-0-0-0);
	\draw[->] (root-0-0-0) -- node [sloped]{$N_{(0,0,0)}^1$} (root-0-0-0-1);
	\draw[->] (root-0-0-1) -- node [sloped,below]{$N_{(0,0,1)}^0$} (root-0-0-1-0);
	\draw[->] (root-0-0-1) -- node [sloped]{$N_{(0,0,1)}^1$} (root-0-0-1-1);
	\draw[->] (root-0-1-0) -- node [sloped,below]{$N_{(0,1,0)}^0$} (root-0-1-0-0);
	\draw[->] (root-0-1-0) -- node [sloped]{$N_{(0,1,0)}^1$} (root-0-1-0-1);
	\draw[->] (root-0-1-1) -- node [sloped,below]{$N_{(0,1,1)}^0$} (root-0-1-1-0);
	\draw[->] (root-0-1-1) -- node [sloped]{$N_{(0,1,1)}^1$} (root-0-1-1-1);
	\draw[->] (root-1-0-0) -- node [sloped,below]{$N_{(1,0,0)}^0$} (root-1-0-0-0);
	\draw[->] (root-1-0-0) -- node [sloped]{$N_{(1,0,0)}^1$} (root-1-0-0-1);
	\draw[->] (root-1-0-1) -- node [sloped,below]{$N_{(1,0,1)}^0$} (root-1-0-1-0);
	\draw[->] (root-1-0-1) -- node [sloped]{$N_{(1,0,1)}^1$} (root-1-0-1-1);
	\draw[->] (root-1-1-0) -- node [sloped,below]{$N_{(1,1,0)}^0$} (root-1-1-0-0);
	\draw[->] (root-1-1-0) -- node [sloped]{$N_{(1,1,0)}^1$} (root-1-1-0-1);
	\draw[->] (root-1-1-1) -- node [sloped,below]{$N_{(1,1,1)}^0$} (root-1-1-1-0);
	\draw[->] (root-1-1-1) -- node [sloped]{$N_{(1,1,1)}^1$} (root-1-1-1-1);
\end{tikzpicture}
}
 \caption{}
        \label{fig:event}
    \end{subfigure}
    \hfill
    \begin{subfigure}{0.49\textwidth}
        \centering
        \resizebox{0.9\linewidth}{!}{%
        \begin{tikzpicture}[auto, scale=14,
	X1_NA/.style={circle,inner sep=1mm,minimum size=0.4cm,draw,black,very thick,fill=white,text=black},
	X2_1/.style={circle,inner sep=1mm,minimum size=0.4cm,draw,black,very thick,fill=white,text=black},
	X3_1/.style={circle,inner sep=1mm,minimum size=0.4cm,draw,black,very thick,fill=yellow,text=black},
    X3_2/.style={circle,inner sep=1mm,minimum size=0.4cm,draw,black,very thick,fill=pink,text=black},
    X3_3/.style={circle,inner sep=1mm,minimum size=0.4cm,draw,black,very thick,fill=green,text=black},
	X4_1/.style={circle,inner sep=1mm,minimum size=0.4cm,draw,black,very thick,fill=orange,text=black},
    X4_2/.style={circle,inner sep=1mm,minimum size=0.4cm,draw,black,very thick,fill=cyan,text=black},
	 leaf/.style={circle,inner sep=1mm,minimum size=0.2cm,draw,very thick,black,fill=gray,text=black}]

	\node [X1_NA] (root) at (0.000000, 0.500000)	{};
	\node [X2_1] (root-0) at (0.250000, 0.233333)	{};
	\node [X2_1] (root-1) at (0.250000, 0.766667)	{};
	\node [X3_1] (root-0-0) at (0.500000, 0.100000)	{};
	\node [X3_2] (root-0-1) at (0.500000, 0.366667)	{};
	\node [X3_3] (root-1-0) at (0.500000, 0.633333)	{};
	\node [X3_3] (root-1-1) at (0.500000, 0.900000)	{};
	\node [X4_1] (root-0-0-0) at (0.750000, 0.033333)	{};
	\node [X4_2] (root-0-0-1) at (0.750000, 0.166667)	{};
	\node [X4_2] (root-0-1-0) at (0.750000, 0.300000)	{};
	\node [X4_1] (root-0-1-1) at (0.750000, 0.433333)	{};
	\node [X4_1] (root-1-0-0) at (0.750000, 0.566667)	{};
	\node [X4_2] (root-1-0-1) at (0.750000, 0.700000)	{};
	\node [X4_2] (root-1-1-0) at (0.750000, 0.833333)	{};
	\node [X4_1] (root-1-1-1) at (0.750000, 0.966667)	{};
	\node [leaf] (root-0-0-0-0) at (1.000000, 0.000000)	{};
	\node [leaf] (root-0-0-0-1) at (1.000000, 0.066667)	{};
	\node [leaf] (root-0-0-1-0) at (1.000000, 0.133333)	{};
	\node [leaf] (root-0-0-1-1) at (1.000000, 0.200000)	{};
	\node [leaf] (root-0-1-0-0) at (1.000000, 0.266667)	{};
	\node [leaf] (root-0-1-0-1) at (1.000000, 0.333333)	{};
	\node [leaf] (root-0-1-1-0) at (1.000000, 0.400000)	{};
	\node [leaf] (root-0-1-1-1) at (1.000000, 0.466667)	{};
	\node [leaf] (root-1-0-0-0) at (1.000000, 0.533333)	{};
	\node [leaf] (root-1-0-0-1) at (1.000000, 0.600000)	{};
	\node [leaf] (root-1-0-1-0) at (1.000000, 0.666667)	{};
	\node [leaf] (root-1-0-1-1) at (1.000000, 0.733333)	{};
	\node [leaf] (root-1-1-0-0) at (1.000000, 0.800000)	{};
	\node [leaf] (root-1-1-0-1) at (1.000000, 0.866667)	{};
	\node [leaf] (root-1-1-1-0) at (1.000000, 0.933333)	{};
	\node [leaf] (root-1-1-1-1) at (1.000000, 1.000000)	{};

	\draw[->] (root) -- node [sloped,below]{$N^0$} (root-0);
	\draw[->] (root) -- node [sloped]{$N^1$} (root-1);
	\draw[->] (root-0) -- node [sloped,below]{$N_{(0)}^0$} (root-0-0);
	\draw[->] (root-0) -- node [sloped]{$N_{(0)}^1$} (root-0-1);
	\draw[->] (root-1) -- node [sloped,below]{$N_{(1)}^0$} (root-1-0);
	\draw[->] (root-1) -- node [sloped]{$N_{(1)}^1$} (root-1-1);
	\draw[->] (root-0-0) -- node [sloped,below]{$N_{(0,0)}^0$} (root-0-0-0);
	\draw[->] (root-0-0) -- node [sloped]{$N_{(0,0)}^1$} (root-0-0-1);
	\draw[->] (root-0-1) -- node [sloped,below]{$N_{(0,1)}^0$} (root-0-1-0);
	\draw[->] (root-0-1) -- node [sloped]{$N_{(0,1)}^1$} (root-0-1-1);
	\draw[->] (root-1-0) -- node [sloped,below]{$N_{(1,0)}^0$} (root-1-0-0);
	\draw[->] (root-1-0) -- node [sloped]{$N_{(1,0)}^1$} (root-1-0-1);
	\draw[->] (root-1-1) -- node [sloped,below]{$N_{(1,1)}^0$} (root-1-1-0);
	\draw[->] (root-1-1) -- node [sloped]{$N_{(1,1)}^1$} (root-1-1-1);
	\draw[->] (root-0-0-0) -- node [sloped,below]{$N_{(0,0,0)}^0$} (root-0-0-0-0);
	\draw[->] (root-0-0-0) -- node [sloped]{$N_{(0,0,0)}^1$} (root-0-0-0-1);
	\draw[->] (root-0-0-1) -- node [sloped,below]{$N_{(0,0,1)}^0$} (root-0-0-1-0);
	\draw[->] (root-0-0-1) -- node [sloped]{$N_{(0,0,1)}^1$} (root-0-0-1-1);
	\draw[->] (root-0-1-0) -- node [sloped,below]{$N_{(0,1,0)}^0$} (root-0-1-0-0);
	\draw[->] (root-0-1-0) -- node [sloped]{$N_{(0,1,0)}^1$} (root-0-1-0-1);
	\draw[->] (root-0-1-1) -- node [sloped,below]{$N_{(0,1,1)}^0$} (root-0-1-1-0);
	\draw[->] (root-0-1-1) -- node [sloped]{$N_{(0,1,1)}^1$} (root-0-1-1-1);
	\draw[->] (root-1-0-0) -- node [sloped,below]{$N_{(1,0,0)}^0$} (root-1-0-0-0);
	\draw[->] (root-1-0-0) -- node [sloped]{$N_{(1,0,0)}^1$} (root-1-0-0-1);
	\draw[->] (root-1-0-1) -- node [sloped,below]{$N_{(1,0,1)}^0$} (root-1-0-1-0);
	\draw[->] (root-1-0-1) -- node [sloped]{$N_{(1,0,1)}^1$} (root-1-0-1-1);
	\draw[->] (root-1-1-0) -- node [sloped,below]{$N_{(1,1,0)}^0$} (root-1-1-0-0);
	\draw[->] (root-1-1-0) -- node [sloped]{$N_{(1,1,0)}^1$} (root-1-1-0-1);
	\draw[->] (root-1-1-1) -- node [sloped,below]{$N_{(1,1,1)}^0$} (root-1-1-1-0);
	\draw[->] (root-1-1-1) -- node [sloped]{$N_{(1,1,1)}^1$} (root-1-1-1-1);
\end{tikzpicture}
        }
        \caption{}
        \label{fig:stage}
    \end{subfigure}
    \caption{Event tree (left) and staged tree (right) for four binary random variables. Edges in the event tree are labeled with counts $N_{\bm{x}{[i-1]}}^{x_i}$ from $\mathcal{D}$. The staged tree is based on the same event tree with $C=\{2,3\}$; vertices with the same color at depth $i$ indicate equal conditional distributions.}
    \label{fig:event-stage}
\end{figure}
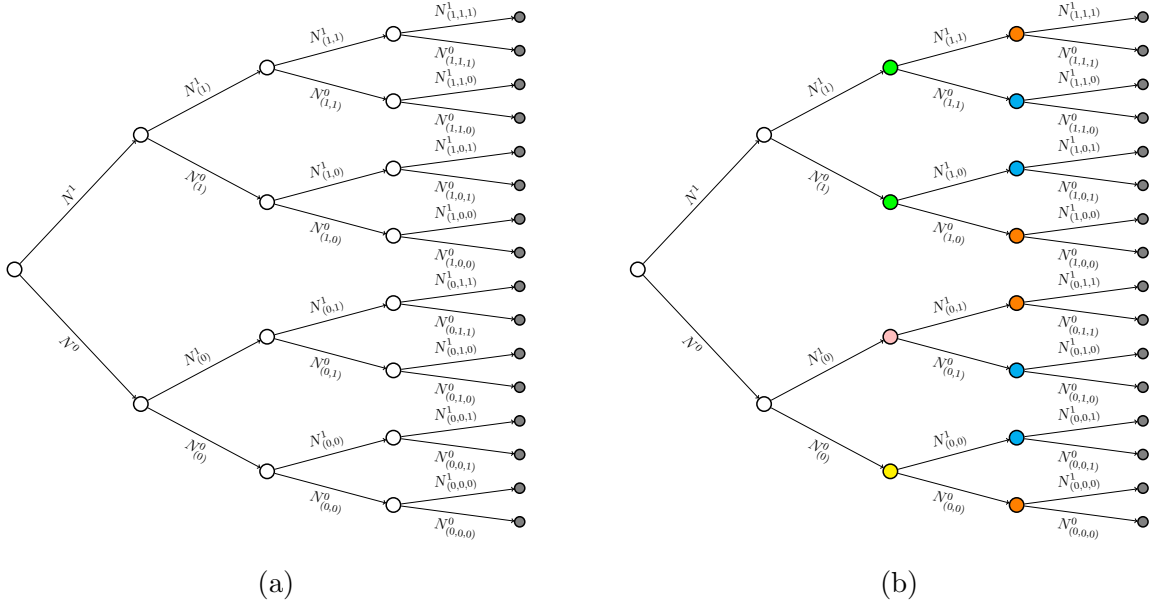

Consider $n$ observations $\mathcal{D} = \{ \bm{x}^{(1)}, \dots, \bm{x}^{(n)} \}$ from $\bm{X}$ where each $\bm{x}^{(k)} = (x_j^{(k)})_{j\in[p]} \in \mathbb{X}$, for $k = 1, \dots, n$. The data can be summarized by $p$ conditional frequency tables, one for each variable $X_i$, $i > 0$. Each table has a row for each $\bm{x}_{[i-1]} \in \mathbb{X}_{[i-1]}$ and a column for each $x_i \in \mathbb{X}_i$. A generic entry is given by
\[
N_{\bm{x}_{[i-1]}}^{x_i} = \sum_{k=1}^n \mathbbm{1}(\bm{x}^{(k)}_{[i-1]} = \bm{x}_{[i-1]}, x_i^{(k)} = x_i),
\]
where $\mathbbm{1}(\cdot)$ is the indicator function. We let $\bm{N}_{\bm{x}_{[i-1]}} = (N_{\bm{x}_{[i-1]}}^{x_i})_{x_i \in \mathbb{X}_i}$ denote the full count vector. For $i = 0$, we simply write $N^{x_0}$. This representation can be visualized using an \emph{event tree} $\mathcal{T}_{\bm{X}}$, where each non-leaf node at depth $i$ corresponds to a context $\bm{x}_{[i-1]} \in \mathbb{X}_{[i-1]}$, and the outgoing edges represent the possible values of $X_i$. Each edge is labeled with the associated count $N_{\bm{x}_{[i-1]}}^{x_i}$. Figure~\ref{fig:stage} shows an example involving four binary variables with $\mathbb{X}_i = \{0, 1\}$ for all $i$.

We now partition the variables into $\bm{X}_C$ and $\bm{X}_{-C}$, where $C = \{c, \dots, p\} \subseteq [p]$ with $c \geq 0$. We interpret $\bm{X}_C$ as categorical variables to be modeled, and $\bm{X}_{-C}$ as categorical covariates. For each $i \in C$, the corresponding table includes one conditional distribution
\[
X_i \mid \bm{X}_{[i-1]} = \bm{x}_{[i-1]} \sim \text{Multinomial}(|\bm{N}_{\bm{x}_{[i-1]}}|, \bm{\theta}_{\bm{x}_{[i-1]}}),
\]
where $\bm{\theta}_{\bm{x}_{[i-1]}}$ is the vector of probabilities over $\mathbb{X}_i$ conditional on $\bm{X}_{[i-1]} = \bm{x}_{[i-1]}$, and  $|\cdot|$ indicates the sum over vector components (e.g., $|\bm{a}_S| = \sum_k a_{S,k}$). Our goal is to identify a partition $\rho_i = \{S_1, \dots, S_{M_i}\}$ of $\mathbb{X}_{[i-1]}$, where $M_i$ denotes the number of node clusters for variable $X_i$, such that $\bm{\theta}_{\bm{x}_{[i-1]}} \equiv \bm{\theta}_{\bm{x}_{[i-1]}'}$ whenever $\bm{x}_{[i-1]}, \bm{x}_{[i-1]}' \in S_m$ for some $m$. These partitions can be visualized as a coloring of the vertices at depth $i$ in the event tree: two vertices receive the same color if their corresponding conditional distributions are equal. An event tree equipped with such a vertex partition is known as a \emph{staged tree}, and each partition block is referred to as a \emph{stage}~\citep{smith2008conditional, collazo2018chain}. Figure~\ref{fig:stage} shows an example of a staged tree based on the event tree in Figure~\ref{fig:event}, with $C = \{2, 3\}$.  For instance, the green stage at depth two encodes the equality $\bm{\theta}_{(1,1)} = \bm{\theta}_{(1,0)}$. We denote a staged tree as the pair $\mathcal{T}_{\bm{X}}^{\bm{\rho}_C} = (\mathcal{T}_{\bm{X}}, \bm{\rho}_C)$, where $\bm{\rho}_C = (\rho_i)_{i \in C}$ contains the partitions for all variables of interest.

\subsection{Staged Tree Models and Conditional Independence}

In DAG-based graphical models, the Markov property provides a direct correspondence between the graph structure and the set of conditional independence statements implied by the underlying distribution~\citep{lauritzen1996graphical}. In staged trees, a similar role is played by the coloring of the vertices: the independence structure of the model is encoded in the partitioning of the tree's vertices into stages. A stage at depth $i$ in the event tree corresponds to a subset $S \subseteq \mathbb{X}_{[i-1]}$ of contexts that share the same conditional distribution for $X_i$. That is, for any $\bm{x}_{[i-1]}, \bm{x}_{[i-1]}' \in S\in\rho_i$, we have
\[
P(X_i \mid \bm{X}_{[i-1]} = \bm{x}_{[i-1]}) = P(X_i \mid \bm{X}_{[i-1]} = \bm{x}_{[i-1]}').
\]
Let $s_i(\bm{x}_{[i-1]})$ denote the stage label at depth $i$. Then the staged tree encodes $X_i \independent X_j \mid \bm{X}_K$ (with $j<i$, $K\subseteq [i-1]\setminus\{j\}$) if and only if $s_i(\bm{x}_{[i-1]})$ is invariant in $x_j$ for fixed $\bm{x}_K$. In other words, if the staging at depth $i$ groups together all contexts that differ only in $x_j$, then $X_i$ is conditionally independent of $X_j$ given $\bm{X}_K$. In Figure~\ref{fig:stage}, at depth $3$, matching colors across contexts $(0,i,j)$ and $(1,i,j)$ (for $i,j\in\{0,1\}$) imply $X_3 \independent X_0 \mid X_1,X_2$.

More flexible patterns of conditional independence can also be represented in staged trees, including several non-symmetric forms~\citep{pensar2016role}. The most widely studied of these is \emph{context-specific conditional independence}~\citep{boutilier1996context}, which refers to independencies that hold only in specific regions of the conditioning space. These types of dependencies cannot be explicitly and graphically represented in DAGs, as they are typically hidden within the structure of the conditional probability tables. In contrast, staged trees make such dependencies visible through their vertex colorings. Formally, for some $j < i$ and $K \subseteq [i-1] \setminus \{j\}$, we say that $X_i$ is context-specifically independent of $X_j$ given a particular context $\bm{X}_K = \bm{x}_K$ if
\[
P(X_i \mid X_j = x_j, \bm{X}_K = \bm{x}_K) = P(X_i \mid \bm{X}_K = \bm{x}_K), \quad \text{for all } x_j \in \mathbb{X}_j.
\]
Equivalently, for fixed $\bm{x}_K$, the stage label $s_i(\bm{x}_{[i-1]})$ does not vary with $x_j$. In other words, within that context, the conditional distribution of $X_i$ is unchanged across values of $X_j$. In Figure~\ref{fig:stage}, at depth $2$, the contexts $(1,0)$ and $(1,1)$ share a color while those with $X_0=0$ do not, encoding $X_2 \independent X_1 \mid X_0=1$.

Beyond symmetric and context-specific independence, staged trees are able to encode non-symmetric patterns such as \emph{partial} and \emph{local} independence~\citep{pensar2016role,varando2024staged}. These are naturally expressed via vertex colorings but are typically less interpretable in complex models, so we do not emphasize them here.

\subsection{Causal Inference with Staged Trees}

A central goal in causal inference is to quantify the effect of a treatment variable on an outcome of interest. Let $T$ be a binary treatment, $Y$ a binary outcome, and $\bm{Z}$ categorical covariates. We model $\bm{X}=(\bm{Z},T,Y)$ with a staged tree in which stages are defined for $T$ and $Y$, while $\bm{Z}$ sets the context and is not clustered. This captures context-specific dependence for treatment and outcome while preserving a fixed frame for adjustment. The primary estimand is the \emph{average treatment effect} (ATE). Using the $\operatorname{do}(\cdot)$ operator~\citep{pearl2009causal}, for binary $T,Y$
\begin{equation}
\label{eq:ate}
\text{ATE}=\mathbb{E}[Y \mid \operatorname{do}(T = 1)]-\mathbb{E}[Y \mid \operatorname{do}(T = 0)].
\end{equation}
The \emph{conditional} ATE (CATE) at covariate profile $\bm{z}$ is the difference between the two interventional expectations in Eq.~\eqref{eq:ate} evaluated at $\bm{Z}=\bm{z}$; it captures treatment-effect heterogeneity. In observational data, under consistency, positivity, and conditional exchangeability~\citep{hernancausal}, the ATE is identifiable by standardization:
\begin{equation}
\label{eq:standardization}
\text{ATE} = \sum_{\bm{z} \in \mathbb{X}_{\bm{Z}}} \left\{ \mathbb{E}[Y \mid T=1, \bm{Z}=\bm{z}] - \mathbb{E}[Y \mid T=0, \bm{Z}=\bm{z}] \right\} P(\bm{Z} = \bm{z}),
\end{equation}
where $P(\bm{Z}=\bm{z})$ is the target-population distribution of covariates. The bracketed term is the CATE at $\bm{z}$.

In staged trees, an intervention $\operatorname{do}(T=t_0)$ is implemented by replacing $P(T\mid\bm{Z})$ with a point mass at $t_0$ and leaving all other factors unchanged, yielding a \emph{causal staged tree}~\citep{leonelli2023context}. The ATE is then computed via Eq.~\eqref{eq:standardization} using probabilities from the interventional tree; CATE values are obtained analogously for each $\bm{z}$. This mirrors classical adjustment and yields consistent estimators under standard assumptions~\citep{varando2025causal}. In the discussion so far, we have assumed that the staged tree structure is known. In most practical settings, however, the true causal model is not observed and must be learned from data. A standard approach would estimate a single model and then compute causal quantities such as the ATE or CATE conditional on that model. However, such post-hoc inference fails to account for model uncertainty and may lead to biased or overconfident conclusions~\citep{berk2013valid}. In the next section, we formalize a fully Bayesian approach that avoids these limitations by jointly estimating the staged tree structure, its parameters, and causal effects within a unified probabilistic framework.

\section{Modeling of Staged Trees via Product Partition Priors}
\label{sec:model}

We adopt the standard Bayesian framework for learning graphical models~\citep{scutari2019learns}, where the goal is to infer a posterior distribution over model structures and their parameters. In our case, the model structure is the staged tree $\mathcal{T}_{\bm{X}}^{\bm{\rho}_C}$, and our primary object of interest is its posterior distribution given the observed data $\mathcal{D}$. Using Bayes' theorem, we can write this in log scale as
\begin{equation*}
    \log P(\mathcal{T}_{\bm{X}}^{\bm{\rho}_C} \mid \mathcal{D}) \propto 
    \log P(\mathcal{D} \mid \mathcal{T}_{\bm{X}}^{\bm{\rho}_C}) + 
    \log P(\mathcal{T}_{\bm{X}}^{\bm{\rho}_C}),
\end{equation*}
where $P(\mathcal{D} \mid \mathcal{T}_{\bm{X}}^{\bm{\rho}_C})$ is the marginal likelihood and $P(\mathcal{T}_{\bm{X}}^{\bm{\rho}_C})$ is the prior distribution over staged trees. The marginal likelihood can be expressed by integrating over the parameter space:
\begin{equation}
\label{eq:marginal}
    P(\mathcal{D} \mid \mathcal{T}_{\bm{X}}^{\bm{\rho}_C}) 
    = \int P(\mathcal{D} \mid \bm{\theta}, \mathcal{T}_{\bm{X}}^{\bm{\rho}_C}) 
    \, P(\bm{\theta} \mid \mathcal{T}_{\bm{X}}^{\bm{\rho}_C}) \, d\bm{\theta},
\end{equation}
where $\bm{\theta} = (\bm{\theta}_i)_{i \in C}$ collects the probability parameters associated with each stage, and $\bm{\theta}_i = (\bm{\theta}_{S_1}, \dots, \bm{\theta}_{S_{M_i}})$ corresponds to the set of multinomial distributions indexed by the partition $\rho_i$. Here, $P(\mathcal{D} \mid \bm{\theta}, \mathcal{T}_{\bm{X}}^{\bm{\rho}_C})$ denotes the likelihood of the data given the staged tree and its parameters, while $P(\bm{\theta} \mid \mathcal{T}_{\bm{X}}^{\bm{\rho}_C})$ defines the prior over stage-specific probabilities under the given structure.

\subsection{The Marginal Likelihood}

As in the case of DAGs, the marginal likelihood in Equation~\eqref{eq:marginal} admits a closed-form expression under standard assumptions~\citep{freeman2011bayesian}. Specifically, if $\mathcal{D}$ is a complete random sample and each stage-specific probability vector $\bm\theta_{S}$ is assigned an independent Dirichlet prior with hyperparameter vector $\bm{a}_S$, then the marginal likelihood can be decomposed as
\begin{equation*}
    \log P(\mathcal{D} \mid \mathcal{T}_{\bm{X}}^{\bm{\rho}_C}) = \sum_{i \in C} \sum_{S \in \rho_i} \log m(\bm{N}_S),
\end{equation*}
where $\bm{N}_S = \sum_{\bm{x}_{[i-1]} \in S} \bm{N}_{\bm{x}_{[i-1]}}$ is the aggregated count vector for stage $S$, associated with variable $X_i$. The function $m(\bm{N}_S)$ corresponds to the marginal likelihood contribution from each stage and takes the form \citep{freeman2011bayesian}:
\begin{equation}
\label{eq:marginal1}
    \log m(\bm{N}_S) = 
    \log \Gamma(|\bm{a}_S|) 
    - \log \Gamma(|\bm{a}_S + \bm{N}_S|) 
    + |\log \Gamma(\bm{a}_S + \bm{N}_S)| 
    - |\log \Gamma(\bm{a}_S)|,
\end{equation}
where $\Gamma(\cdot)$ denotes the Gamma function. In line with standard practice for graphical models, we assume a symmetric Dirichlet prior by setting each entry of $\bm{a}_S$ to $a / \#\mathbb{X}_i$, for some $a > 0$, though other choices of hyperparameters are possible.

\subsection{The Prior over Staged Trees}

The final component of the proposed model is the prior distribution over the space of staged trees, which corresponds to a prior over the space of vertex partitions $\bm\rho_C$. For notational simplicity, we write $P(\mathcal{T}_{\bm{X}}^{\bm{\rho}_C}) \equiv P(\bm{\rho}_C)$. To reduce complexity, we assume \emph{structure modularity}~\citep{Koller}, so that the prior factorizes over the variables of interest:
\begin{equation}
\label{eq:priorfact}
P(\bm{\rho}_C) = \prod_{i \in C} P(\rho_i).
\end{equation}
This assumption implies that the stage partitions at different depths of the tree are a priori independent, and is standard in the Bayesian literature on graphical model learning. See Section~\ref{sec:concl} for further discussion on its implications in the context of staged trees. With the exception of~\citet{collazo2016new}, who proposed a non-local prior to penalize excessive merging, most existing approaches assume $P(\rho_i)$ to be uniform over the space of partitions. However, it is well-documented that such uniform priors tend to favor overly complex structures, leading to models that lack parsimony and interpretability~\citep{collazo2016new, eggeling2019structure}. Given the close connection between learning a staged tree and clustering the contexts at each level of the tree, it is natural to consider a prior grounded in the framework of \emph{product partition models} (PPMs)~\citep{quintana2003bayesian}. In this class of models, the prior over partitions is defined in terms of a \emph{cohesion function} $c(S)$ for each block $S$ in the partition, which quantifies the prior belief that the elements of $S$ should be grouped together. The induced prior on a partition $\rho_i = \{S_1, \dots, S_{M_i}\}$ takes the form
\begin{equation*}
\label{eq:ppm}
P(\rho_i = \{S_1, \dots, S_{M_i}\}) \propto \prod_{j=1}^{M_i} c(S_j).
\end{equation*}
A popular choice of cohesion function is $c(S_j) = \kappa \cdot \Gamma(\#S_j)$, where $\kappa > 0$ controls the expected number of clusters and leads to the exchangeable partition probability function (eppf) of the Dirichlet process~\citep{antoniak1974mixtures}. Larger values of $\kappa$ encourage more stages, while smaller values promote simpler, more parsimonious trees~\citep{muller2011product}.

\subsubsection{Distance-penalized product partition priors}

\begin{figure}
    \centering 
    \begin{subfigure}[b]{0.6\textwidth}
   \centering
    \begin{tikzpicture}[auto, scale=14,
	X1_NA/.style={circle,inner sep=1mm,minimum size=0.4cm,draw,black,very thick,fill=white,text=black},
	X2_1/.style={circle,inner sep=1mm,minimum size=0.4cm,draw,black,very thick,fill=white,text=black},
	X3_1/.style={circle,inner sep=1mm,minimum size=0.4cm,draw,black,very thick,fill=yellow,text=black},
    X3_2/.style={circle,inner sep=1mm,minimum size=0.4cm,draw,black,very thick,fill=pink,text=black},
    X3_3/.style={circle,inner sep=1mm,minimum size=0.4cm,draw,black,very thick,fill=green,text=black},
	X4_1/.style={circle,inner sep=1mm,minimum size=0.4cm,draw,black,very thick,fill=orange,text=black},
    X4_2/.style={circle,inner sep=1mm,minimum size=0.4cm,draw,black,very thick,fill=cyan,text=black},
	 leaf/.style={circle,inner sep=1mm,minimum size=0.2cm,draw,very thick,black,fill=gray,text=black}]

	\node [X1_NA] (root) at (0.000000, 0.500000)	{};
	\node [X2_1] (root-0) at (0.250000, 0.233333)	{};
	\node [X2_1] (root-1) at (0.250000, 0.766667)	{};
	\node [X2_1] (root-0-0) at (0.500000, 0.100000)	{};
	\node [X2_1] (root-0-1) at (0.500000, 0.366667)	{};
	\node [X2_1] (root-1-0) at (0.500000, 0.633333)	{};
	\node [X2_1] (root-1-1) at (0.500000, 0.900000)	{};
	\node [X4_1] (root-0-0-0) at (0.750000, 0.033333)	{};
	\node [X4_2] (root-0-0-1) at (0.750000, 0.166667)	{};
	\node [X2_1] (root-0-1-0) at (0.750000, 0.300000)	{};
	\node [X2_1] (root-0-1-1) at (0.750000, 0.433333)	{};
	\node [X3_1] (root-1-0-0) at (0.750000, 0.566667)	{};
	\node [X2_1] (root-1-0-1) at (0.750000, 0.700000)	{};
	\node [X2_1] (root-1-1-0) at (0.750000, 0.833333)	{};
	\node [X3_3] (root-1-1-1) at (0.750000, 0.966667)	{};
	\node [leaf] (root-0-0-0-0) at (1.000000, 0.000000)	{};
	\node [leaf] (root-0-0-0-1) at (1.000000, 0.066667)	{};
	\node [leaf] (root-0-0-1-0) at (1.000000, 0.133333)	{};
	\node [leaf] (root-0-0-1-1) at (1.000000, 0.200000)	{};
	\node [leaf] (root-0-1-0-0) at (1.000000, 0.266667)	{};
	\node [leaf] (root-0-1-0-1) at (1.000000, 0.333333)	{};
	\node [leaf] (root-0-1-1-0) at (1.000000, 0.400000)	{};
	\node [leaf] (root-0-1-1-1) at (1.000000, 0.466667)	{};
	\node [leaf] (root-1-0-0-0) at (1.000000, 0.533333)	{};
	\node [leaf] (root-1-0-0-1) at (1.000000, 0.600000)	{};
	\node [leaf] (root-1-0-1-0) at (1.000000, 0.666667)	{};
	\node [leaf] (root-1-0-1-1) at (1.000000, 0.733333)	{};
	\node [leaf] (root-1-1-0-0) at (1.000000, 0.800000)	{};
	\node [leaf] (root-1-1-0-1) at (1.000000, 0.866667)	{};
	\node [leaf] (root-1-1-1-0) at (1.000000, 0.933333)	{};
	\node [leaf] (root-1-1-1-1) at (1.000000, 1.000000)	{};

	\draw[->] (root) -- node [sloped,below]{$N^0$} (root-0);
	\draw[->] (root) -- node [sloped]{$N^1$} (root-1);
	\draw[->] (root-0) -- node [sloped,below]{$N_{(0)}^0$} (root-0-0);
	\draw[->] (root-0) -- node [sloped]{$N_{(0)}^1$} (root-0-1);
	\draw[->] (root-1) -- node [sloped,below]{$N_{(1)}^0$} (root-1-0);
	\draw[->] (root-1) -- node [sloped]{$N_{(1)}^1$} (root-1-1);
	\draw[->] (root-0-0) -- node [sloped,below]{$N_{(0,0)}^0$} (root-0-0-0);
	\draw[->] (root-0-0) -- node [sloped]{$N_{(0,0)}^1$} (root-0-0-1);
	\draw[->] (root-0-1) -- node [sloped,below]{$N_{(0,1)}^0$} (root-0-1-0);
	\draw[->] (root-0-1) -- node [sloped]{$N_{(0,1)}^1$} (root-0-1-1);
	\draw[->] (root-1-0) -- node [sloped,below]{$N_{(1,0)}^0$} (root-1-0-0);
	\draw[->] (root-1-0) -- node [sloped]{$N_{(1,0)}^1$} (root-1-0-1);
	\draw[->] (root-1-1) -- node [sloped,below]{$N_{(1,1)}^0$} (root-1-1-0);
	\draw[->] (root-1-1) -- node [sloped]{$N_{(1,1)}^1$} (root-1-1-1);
	\draw[->] (root-0-0-0) -- node [sloped,below]{$N_{(0,0,0)}^0$} (root-0-0-0-0);
	\draw[->] (root-0-0-0) -- node [sloped]{$N_{(0,0,0)}^1$} (root-0-0-0-1);
	\draw[->] (root-0-0-1) -- node [sloped,below]{$N_{(0,0,1)}^0$} (root-0-0-1-0);
	\draw[->] (root-0-0-1) -- node [sloped]{$N_{(0,0,1)}^1$} (root-0-0-1-1);
	\draw[->] (root-0-1-0) -- node [sloped,below]{$N_{(0,1,0)}^0$} (root-0-1-0-0);
	\draw[->] (root-0-1-0) -- node [sloped]{$N_{(0,1,0)}^1$} (root-0-1-0-1);
	\draw[->] (root-0-1-1) -- node [sloped,below]{$N_{(0,1,1)}^0$} (root-0-1-1-0);
	\draw[->] (root-0-1-1) -- node [sloped]{$N_{(0,1,1)}^1$} (root-0-1-1-1);
	\draw[->] (root-1-0-0) -- node [sloped,below]{$N_{(1,0,0)}^0$} (root-1-0-0-0);
	\draw[->] (root-1-0-0) -- node [sloped]{$N_{(1,0,0)}^1$} (root-1-0-0-1);
	\draw[->] (root-1-0-1) -- node [sloped,below]{$N_{(1,0,1)}^0$} (root-1-0-1-0);
	\draw[->] (root-1-0-1) -- node [sloped]{$N_{(1,0,1)}^1$} (root-1-0-1-1);
	\draw[->] (root-1-1-0) -- node [sloped,below]{$N_{(1,1,0)}^0$} (root-1-1-0-0);
	\draw[->] (root-1-1-0) -- node [sloped]{$N_{(1,1,0)}^1$} (root-1-1-0-1);
	\draw[->] (root-1-1-1) -- node [sloped,below]{$N_{(1,1,1)}^0$} (root-1-1-1-0);
	\draw[->] (root-1-1-1) -- node [sloped]{$N_{(1,1,1)}^1$} (root-1-1-1-1);
\end{tikzpicture}
    \caption{}
    \end{subfigure}%
    \hfill
    \begin{subfigure}[b]{0.35\textwidth}
    \centering
    \vfill
    \begin{minipage}{\textwidth}
        \centering
        \begin{tabular}{lc}
            \toprule
            \textbf{Vertices} & \textbf{Distance} \\
            \midrule
            \colorcircle{orange}\colorcircle{cyan} & 0.33 \\
            \colorcircle{orange}\colorcircle{yellow}& 0.33\\
            \colorcircle{orange}\colorcircle{green}& 1 \\
            \colorcircle{cyan}\colorcircle{yellow}& 0.66\\
            \colorcircle{cyan}\colorcircle{green} & 0.66\\
            \colorcircle{yellow}\colorcircle{green}& 0.66 \\
            \bottomrule
        \end{tabular}
    \end{minipage}
    \vfill
    \caption{}
    \end{subfigure}
    \caption{(a) An example of an event tree and (b) the normalized tree-based Hamming distance between some of its vertices.}
    \label{fig:tree_and_table}
\end{figure}
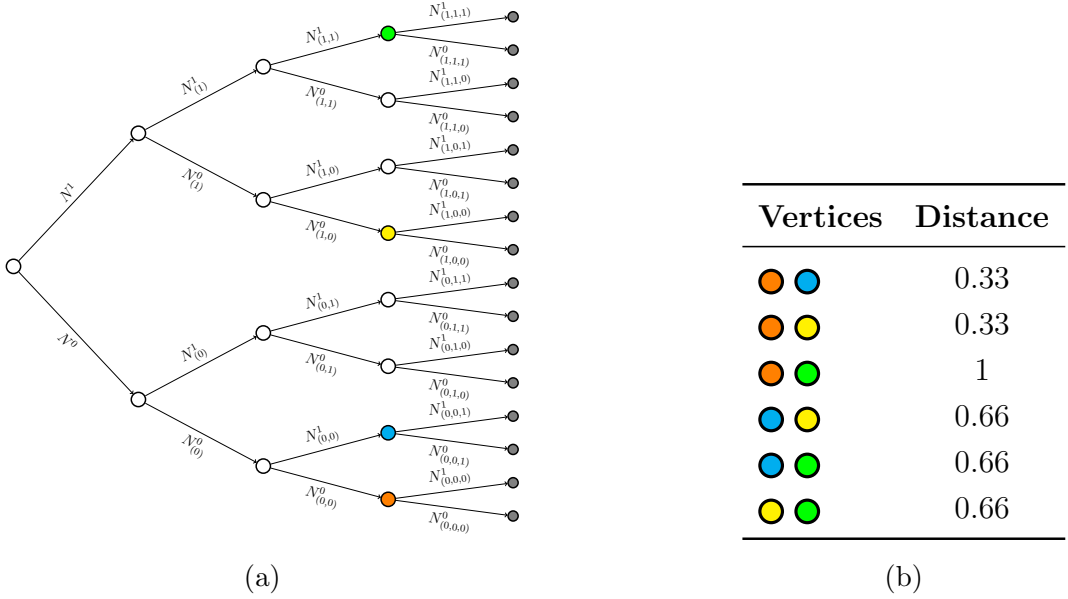

Unlike standard clustering tasks where the objects being grouped are independent and exchangeable, staged tree learning involves clustering the vertices of an event tree, each of which corresponds to a specific context defined by earlier variable assignments. These contexts are not interchangeable: they carry semantic meaning and occupy a well-defined position in the tree structure. As a result, it is natural to incorporate information about their similarity when deciding how to group them into stages. Recent developments in PPMs have focused on incorporating prior knowledge into the clustering process, including covariate-based or spatially structured penalties~\citep{hegarty2008bayesian, muller2011product, page2016spatial}. Inspired by the prior formulation proposed by~\citet{cremaschi2023change}, we define a prior over $P(\rho_i)$ that incorporates pairwise distances between vertices. This formulation encourages parsimonious partitions while favoring the grouping of similar contexts. Formally, we consider the following eppf:
\begin{equation}
\label{eq:prior}
P(\rho_i = \{S_1, \dots, S_{M_i}\}) \propto 
\kappa^{M_i} \prod_{j=1}^{M_i} \Gamma(\# S_j) 
\exp\left( -\xi \sum_{k, \ell \in S_j} d_{k,\ell} \right),
\end{equation}
where $\xi \geq 0$ controls the strength of the penalty, and $d_{k,\ell}$ is a distance function measuring dissimilarity between contexts $k$ and $\ell$ within the same stage. This prior favors compact partitions when $\xi$ is large, while we recover the Dirichlet process eppf when $\xi = 0$. To define the pairwise distances $d_{k,\ell}$, we introduce the \textit{normalized tree-based Hamming distance}, which compares the configurations associated with two vertices in the tree. For two contexts $\bm{x}_{[i-1]}, \bm{x}_{[i-1]}' \in \mathbb{X}_{[i-1]}$, the distance is defined as
\[
d_{\bm{x}_{[i-1]},\bm{x}_{[i-1]}'} = 1 - \frac{\#\{j \in [i-1] : x_j = x_j'\}}{i-1},
\]
where the numerator counts the number of positions at which the two contexts agree. This distance lies in the open interval $(0,1]$ and reflects how similar the two contexts are, with smaller values indicating greater similarity. Figure~\ref{fig:tree_and_table} illustrates the normalized tree-based Hamming distance between four vertices in a small staged tree. Vertices that are closer under this metric often reflect more interpretable patterns of dependence. For example, the orange–cyan and orange–yellow pairs differ in only one component of their contexts, suggesting \emph{context-specific independence}, a structure well captured by staged trees but not representable in DAGs. In contrast, the orange–green pair has a maximum distance of 1, reflecting more heterogeneous contexts and suggesting a form of \emph{local dependence}, which lacks a systematic independence interpretation. The formulation in Equation~\eqref{eq:prior} is thus designed to favor partitions that group similar contexts, encouraging stage structures that support interpretable inferences. To illustrate the effect of the distance-penalized prior, Supplementary Table S1 
reports its values for all partitions of four vertices at depth two. In summary, increasing $\kappa$ raises the probability of partitions with more blocks, while larger $\xi$ penalizes groupings of dissimilar vertices, showing that the proposed prior favors structurally coherent stage groupings.

To assess the effect of the distance-sensitive prior, we simulate observations ($n \in \{500,1000,2500,5000,7500,10000\}$) from a staged tree (Supplementary Figure S1) 
and estimate the partition of the 32 vertices of the last variable using the MCMC algorithm in Supplementary Section 1. 
Each combination of $\xi \in \{1,1/2,1/4,1/8,1/16,1/32,1/64,0\}$ and $\kappa \in \{0.01,0.05,0.1,0.5,1,5\}$ is run for 3000 iterations, with 1000 burn-in and thinning by two, yielding 1000 posterior samples. Supplementary Figure S2 
shows the median number of estimated stages over five replicates. As expected, the influence of the prior wanes with increasing sample size. We also observe an approximate inverse relationship between $\xi$ and $\kappa$ that yields an “iso-complexity” ridge: increasing one while decreasing the other produces similar model complexity. While fully Bayesian mixing over $(\xi,\kappa)$ is ideal, it is computationally heavy at this scale; in practice we fix them, and results are stable once $n$ is moderately large. For small $n$, different pairs often lead to comparable complexity, so one can fix one hyperparameter and tune the other.

\subsubsection{Incorporating Continuous Covariates via Covariate-Dependent Priors}

In many applications, one may have access to additional continuous variables that are informative about the grouping of contexts but are not of direct interest in the staged tree. These covariates, denoted by 
$\bm{Z} = (Z_{1}, \dots, Z_{q})$, 
are not included as tree variables and may not be suitable for discretization. To incorporate this information, we introduce a covariate-augmented prior over partitions inspired by PPMx \citep{muller2011product}.  We define the prior as:
\begin{equation*}
\label{eq:ppmx}
P(\rho_i = \{S_1, \dots, S_{M_i}\}) \propto 
\kappa^{M_i} \prod_{j=1}^{M_i} \Gamma(\# S_j)
\exp\left( -\xi \sum_{k,\ell \in S_j} d_{k,\ell}
- \sum_{z=1}^{q} \lambda_z \sum_{k,\ell \in S_j} \delta_{k,\ell}^{(z)} \right),
\end{equation*}
where \(\lambda_z \geq 0\) weighs the covariate penalty for each continuous variable \(Z_{z}\), with \(z = 1, \dots, q\). Model \eqref{eq:prior} is obtained when $\lambda_z = 0$. The term \(\delta_{k,\ell}^{(z)}\) captures the dissimilarity between contexts \(k\) and \(\ell\) in covariate \(Z_z\). To define it, we assume a Normal-Inverse-Gamma model for \(Z_z\) with hyperparameters \((m_0, \kappa_0, \alpha_0, \beta_0)\) \citep{murphy2012machine} and compute for each context \(k\)
\[
\delta_{k,\ell}^{(z)} = -\left[
\log p(\bm{z}^{(z)}_{k \cup \ell}) - \log p(\bm{z}^{(z)}_k) - \log p(\bm{z}^{(z)}_\ell)
\right],
\]
where \(\bm{z}^{(z)}_k\) denotes the observed values of covariate \(Z_{z}\) in context \(k\) and, for a set $S$, the marginal likelihood is given by:
\begin{multline*}
\log p(\bm{z}^{(z)}_S) = \alpha_0 \log \beta_0 + \tfrac{1}{2} \log \kappa_0 
- \log \Gamma(\alpha_0) - \tfrac{n_S}{2} \log (2\pi) 
+ \log \Gamma\left(\alpha_0 + \tfrac{n_S}{2}\right) \\
 - \tfrac{1}{2} \log(\kappa_0 + n_S)
- \left(\alpha_0 + \tfrac{n_S}{2}\right) \log\left(\beta_0 + \tfrac{1}{2} s_S^{(z)} + \tfrac{\kappa_0 n_S (\bar{\bm{z}}_S^{(z)} - m_0)^2}{2(\kappa_0 + n_S)}\right),
\end{multline*}
with \(\bar{\bm{z}}_S^{(z)}\) the sample mean, \(s^{(z)}_S\) the total sum of squared deviations from the mean, and \(n_S\) the number of observations in set \(S\). These quantities are computed over the subset of observations falling in context \(S\), using only the values of covariate \(Z_z\). This modified eppf encourages merging contexts that are similar in terms of both tree position and covariate distribution, leading to stage groupings that are structurally and statistically coherent. In practice, the hyperparameters of the Normal-Inverse-Gamma prior can be set to weakly informative values to ensure stability across different contexts. A common choice is to set the prior mean to \( m_0 = 0 \), again assuming standardized covariates, and to fix \( \kappa_0 = 1 \) to give moderate weight to this mean. The shape and scale parameters \( \alpha_0 = \beta_0 = 1 \) define a vague prior over the variance. Standardizing covariates before modeling is recommended to make these default settings broadly applicable.

\begin{figure}
  \centering

  \begin{subfigure}[t]{0.29\textwidth}
    \centering
    \begin{tikzpicture}[auto, scale=12,
	X1_NA/.style={circle,inner sep=1mm,minimum size=0.4cm,draw,black,very thick,fill=white,text=black},
	X2_1/.style={circle,inner sep=1mm,minimum size=0.4cm,draw,black,very thick,fill=orange,text=black},
    X2_2/.style={circle,inner sep=1mm,minimum size=0.4cm,draw,black,very thick,fill=purple,text=black},
	X3_1/.style={circle,inner sep=1mm,minimum size=0.4cm,draw,black,very thick,fill=yellow,text=black},
    X3_2/.style={circle,inner sep=1mm,minimum size=0.4cm,draw,black,very thick,fill=pink,text=black},
    X3_3/.style={circle,inner sep=1mm,minimum size=0.4cm,draw,black,very thick,fill=cyan,text=black},
	X4_1/.style={circle,inner sep=1mm,minimum size=0.4cm,draw,black,very thick,fill=blue,text=black},
    X4_2/.style={circle,inner sep=1mm,minimum size=0.4cm,draw,black,very thick,fill=green,text=black},
    X4_3/.style={circle,inner sep=1mm,minimum size=0.4cm,draw,black,very thick,fill=red,text=black},
	 leaf/.style={circle,inner sep=1mm,minimum size=0.2cm,draw,very thick,black,fill=gray,text=black}]

	\node [X1_NA] (root) at (0.000000, 0.500000)	{};
	\node [X2_1] (root-0) at (0.250000, 0.233333)	{};
	\node [X2_2] (root-1) at (0.250000, 0.766667)	{};
	\node [X3_1] (root-0-0) at (0.500000, 0.100000)	{};
	\node [X3_2] (root-0-1) at (0.500000, 0.366667)	{};
	\node [X3_1] (root-1-0) at (0.500000, 0.633333)	{};
	\node [X3_3] (root-1-1) at (0.500000, 0.900000)	{};
	\node [X4_1] (root-0-0-0) at (0.750000, 0.033333)	{};
	\node [X4_1] (root-0-0-1) at (0.750000, 0.166667)	{};
	\node [X4_2] (root-0-1-0) at (0.750000, 0.300000)	{};
	\node [X4_2] (root-0-1-1) at (0.750000, 0.433333)	{};
	\node [X4_1] (root-1-0-0) at (0.750000, 0.566667)	{};
	\node [X4_1] (root-1-0-1) at (0.750000, 0.700000)	{};
	\node [X4_3] (root-1-1-0) at (0.750000, 0.833333)	{};
	\node [X4_3] (root-1-1-1) at (0.750000, 0.966667)	{};
	\node [leaf] (root-0-0-0-0) at (1.000000, 0.000000)	{};
	\node [leaf] (root-0-0-0-1) at (1.000000, 0.066667)	{};
	\node [leaf] (root-0-0-1-0) at (1.000000, 0.133333)	{};
	\node [leaf] (root-0-0-1-1) at (1.000000, 0.200000)	{};
	\node [leaf] (root-0-1-0-0) at (1.000000, 0.266667)	{};
	\node [leaf] (root-0-1-0-1) at (1.000000, 0.333333)	{};
	\node [leaf] (root-0-1-1-0) at (1.000000, 0.400000)	{};
	\node [leaf] (root-0-1-1-1) at (1.000000, 0.466667)	{};
	\node [leaf] (root-1-0-0-0) at (1.000000, 0.533333)	{};
	\node [leaf] (root-1-0-0-1) at (1.000000, 0.600000)	{};
	\node [leaf] (root-1-0-1-0) at (1.000000, 0.666667)	{};
	\node [leaf] (root-1-0-1-1) at (1.000000, 0.733333)	{};
	\node [leaf] (root-1-1-0-0) at (1.000000, 0.800000)	{};
	\node [leaf] (root-1-1-0-1) at (1.000000, 0.866667)	{};
	\node [leaf] (root-1-1-1-0) at (1.000000, 0.933333)	{};
	\node [leaf] (root-1-1-1-1) at (1.000000, 1.000000)	{};

	\draw[->] (root) -- node [sloped,below]{\Large$0.60$} (root-0);
	\draw[->] (root) -- node [sloped]{\Large$0.40$} (root-1);
	\draw[->] (root-0) -- node [sloped,below]{\Large$0.70$} (root-0-0);
	\draw[->] (root-0) -- node [sloped]{\Large$0.30$} (root-0-1);
	\draw[->] (root-1) -- node [sloped,below]{\Large$0.30$} (root-1-0);
	\draw[->] (root-1) -- node [sloped]{\Large$0.70$} (root-1-1);
	\draw[->] (root-0-0) -- node [sloped,below]{\Large$0.40$} (root-0-0-0);
	\draw[->] (root-0-0) -- node [sloped]{\Large$0.60$} (root-0-0-1);
	\draw[->] (root-0-1) -- node [sloped,below]{\Large$0.70$} (root-0-1-0);
	\draw[->] (root-0-1) -- node [sloped]{\Large$0.30$} (root-0-1-1);
	\draw[->] (root-1-0) -- node [sloped,below]{\Large$0.40$} (root-1-0-0);
	\draw[->] (root-1-0) -- node [sloped]{\Large$0.60$} (root-1-0-1);
	\draw[->] (root-1-1) -- node [sloped,below]{\Large$0.75$} (root-1-1-0);
	\draw[->] (root-1-1) -- node [sloped]{\Large$0.25$} (root-1-1-1);
	\draw[->] (root-0-0-0) -- node [sloped,below]{\Large $0.70$} (root-0-0-0-0);
	\draw[->] (root-0-0-0) -- node [sloped]{\Large$0.30$} (root-0-0-0-1);
	\draw[->] (root-0-0-1) -- node [sloped,below]{\Large$0.70$} (root-0-0-1-0);
	\draw[->] (root-0-0-1) -- node [sloped]{\Large$0.30$} (root-0-0-1-1);
	\draw[->] (root-0-1-0) -- node [sloped,below]{\Large$0.30$} (root-0-1-0-0);
	\draw[->] (root-0-1-0) -- node [sloped]{\Large$0.70$} (root-0-1-0-1);
	\draw[->] (root-0-1-1) -- node [sloped,below]{\Large$0.30$} (root-0-1-1-0);
	\draw[->] (root-0-1-1) -- node [sloped]{\Large$0.70$} (root-0-1-1-1);
	\draw[->] (root-1-0-0) -- node [sloped,below]{\Large$0.70$} (root-1-0-0-0);
	\draw[->] (root-1-0-0) -- node [sloped]{\Large$0.30$} (root-1-0-0-1);
	\draw[->] (root-1-0-1) -- node [sloped,below]{\Large$0.70$} (root-1-0-1-0);
	\draw[->] (root-1-0-1) -- node [sloped]{\Large$0.30$} (root-1-0-1-1);
	\draw[->] (root-1-1-0) -- node [sloped,below]{\Large$0.75$} (root-1-1-0-0);
	\draw[->] (root-1-1-0) -- node [sloped]{\Large$0.25$} (root-1-1-0-1);
	\draw[->] (root-1-1-1) -- node [sloped,below]{\Large$0.75$} (root-1-1-1-0);
	\draw[->] (root-1-1-1) -- node [sloped]{\Large$0.25$} (root-1-1-1-1);
\end{tikzpicture}
    \caption{}
  \end{subfigure}
  \hfill
  \begin{subfigure}[t]{0.29\textwidth}
    \centering
\includegraphics[scale=0.09]{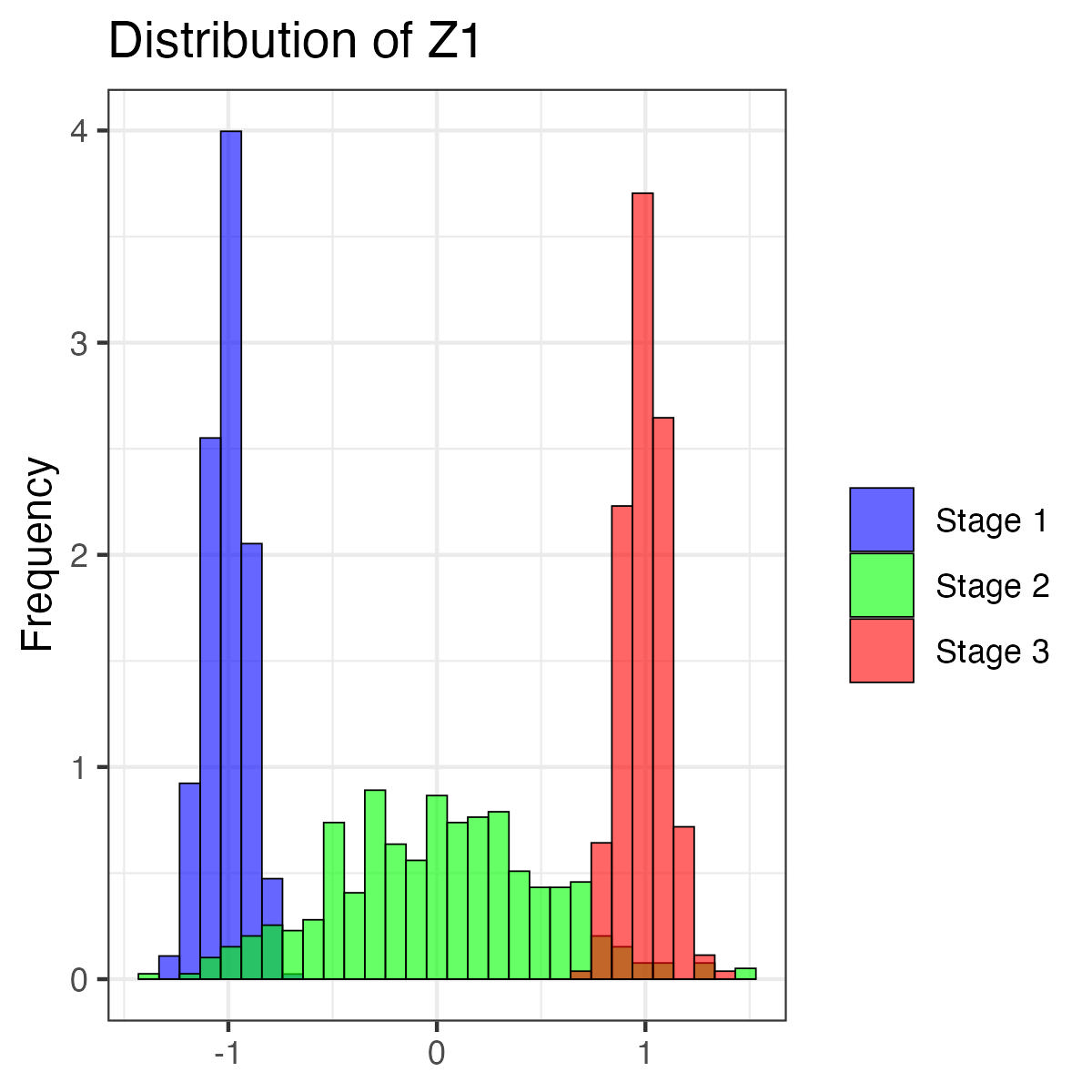}
\caption{}
  \end{subfigure}
  \begin{subfigure}[t]{0.19\textwidth}
    \centering
    \begin{tikzpicture}[auto, scale=12,
	X1_NA/.style={circle,inner sep=1mm,minimum size=0.6cm,draw,black,very thick,fill=white,text=black},
	X2_1/.style={circle,inner sep=1mm,minimum size=0.6cm,draw,black,very thick,fill=orange,text=black},
    X2_2/.style={circle,inner sep=1mm,minimum size=0.6cm,draw,black,very thick,fill=purple,text=black},
	X3_1/.style={circle,inner sep=1mm,minimum size=0.6cm,draw,black,very thick,fill=yellow,text=black},
    X3_2/.style={circle,inner sep=1mm,minimum size=0.6cm,draw,black,very thick,fill=pink,text=black},
    X3_3/.style={circle,inner sep=1mm,minimum size=0.6cm,draw,black,very thick,fill=cyan,text=black},
	X4_1/.style={circle,inner sep=1mm,minimum size=0.6cm,draw,black,very thick,fill=blue,text=black},
    X4_2/.style={circle,inner sep=1mm,minimum size=0.6cm,draw,black,very thick,fill=green,text=black},
    X4_3/.style={circle,inner sep=1mm,minimum size=0.6cm,draw,black,very thick,fill=red,text=black},
	 leaf/.style={circle,inner sep=1mm,minimum size=0.4cm,draw,very thick,black,fill=gray,text=black}]

	\node [X1_NA] (root) at (0.000000, 0.500000*\dd)	{};
	\node [X2_1] (root-0) at (0.2*\ddd, 0.233333*\dd)	{};
	\node [X2_2] (root-1) at (0.2*\ddd, 0.766667*\dd)	{};
	\node [X3_1] (root-0-0) at (0.4*\ddd, 0.100000*\dd)	{};
	\node [X3_2] (root-0-1) at (0.4*\ddd, 0.366667*\dd)	{};
	\node [X3_1] (root-1-0) at (0.4*\ddd, 0.633333*\dd)	{};
	\node [X3_2] (root-1-1) at (0.4*\ddd, 0.900000*\dd)	{};
	\node [X4_1] (root-0-0-0) at (0.6*\ddd, 0.033333*\dd)	{};
	\node [X4_1] (root-0-0-1) at (0.6*\ddd, 0.166667*\dd)	{};
	\node [X4_2] (root-0-1-0) at (0.60000*\ddd, 0.300000*\dd)	{};
	\node [X4_2] (root-0-1-1) at (0.60000*\ddd, 0.433333*\dd)	{};
	\node [X4_1] (root-1-0-0) at (0.60000*\ddd, 0.566667*\dd)	{};
	\node [X4_1] (root-1-0-1) at (0.60000*\ddd, 0.700000*\dd)	{};
	\node [X4_1] (root-1-1-0) at (0.60000*\ddd, 0.833333*\dd)	{};
	\node [X4_1] (root-1-1-1) at (0.60000*\ddd, 0.966667*\dd)	{};
	\node [leaf] (root-0-0-0-0) at (0.8000000*\ddd, 0.000000*\dd)	{};
	\node [leaf] (root-0-0-0-1) at (0.8000000*\ddd, 0.066667*\dd)	{};
	\node [leaf] (root-0-0-1-0) at (0.8000000*\ddd, 0.133333*\dd)	{};
	\node [leaf] (root-0-0-1-1) at (0.8000000*\ddd, 0.200000*\dd)	{};
	\node [leaf] (root-0-1-0-0) at (0.8000000*\ddd, 0.266667*\dd)	{};
	\node [leaf] (root-0-1-0-1) at (0.8000000*\ddd, 0.333333*\dd)	{};
	\node [leaf] (root-0-1-1-0) at (0.8000000*\ddd, 0.400000*\dd)	{};
	\node [leaf] (root-0-1-1-1) at (0.8000000*\ddd, 0.466667*\dd)	{};
	\node [leaf] (root-1-0-0-0) at (0.8000000*\ddd, 0.533333*\dd)	{};
	\node [leaf] (root-1-0-0-1) at (0.8000000*\ddd, 0.600000*\dd)	{};
	\node [leaf] (root-1-0-1-0) at (0.8000000*\ddd, 0.666667*\dd)	{};
	\node [leaf] (root-1-0-1-1) at (0.8000000*\ddd, 0.733333*\dd)	{};
	\node [leaf] (root-1-1-0-0) at (0.8000000*\ddd, 0.800000*\dd)	{};
	\node [leaf] (root-1-1-0-1) at (0.8000000*\ddd, 0.866667*\dd)	{};
	\node [leaf] (root-1-1-1-0) at (0.8000000*\ddd, 0.933333*\dd)	{};
	\node [leaf] (root-1-1-1-1) at (0.8000000*\ddd, 1.000000*\dd)	{};

	\draw[->] (root) -- node [sloped,below]{} (root-0);
	\draw[->] (root) -- node [sloped]{} (root-1);
	\draw[->] (root-0) -- node [sloped,below]{} (root-0-0);
	\draw[->] (root-0) -- node [sloped]{} (root-0-1);
	\draw[->] (root-1) -- node [sloped,below]{} (root-1-0);
	\draw[->] (root-1) -- node [sloped]{} (root-1-1);
	\draw[->] (root-0-0) -- node [sloped,below]{} (root-0-0-0);
	\draw[->] (root-0-0) -- node [sloped]{} (root-0-0-1);
	\draw[->] (root-0-1) -- node [sloped,below]{} (root-0-1-0);
	\draw[->] (root-0-1) -- node [sloped]{} (root-0-1-1);
	\draw[->] (root-1-0) -- node [sloped,below]{} (root-1-0-0);
	\draw[->] (root-1-0) -- node [sloped]{} (root-1-0-1);
	\draw[->] (root-1-1) -- node [sloped,below]{} (root-1-1-0);
	\draw[->] (root-1-1) -- node [sloped]{} (root-1-1-1);
	\draw[->] (root-0-0-0) -- node [sloped,below]{} (root-0-0-0-0);
	\draw[->] (root-0-0-0) -- node [sloped]{} (root-0-0-0-1);
	\draw[->] (root-0-0-1) -- node [sloped,below]{} (root-0-0-1-0);
	\draw[->] (root-0-0-1) -- node [sloped]{} (root-0-0-1-1);
	\draw[->] (root-0-1-0) -- node [sloped,below]{} (root-0-1-0-0);
	\draw[->] (root-0-1-0) -- node [sloped]{} (root-0-1-0-1);
	\draw[->] (root-0-1-1) -- node [sloped,below]{} (root-0-1-1-0);
	\draw[->] (root-0-1-1) -- node [sloped]{} (root-0-1-1-1);
	\draw[->] (root-1-0-0) -- node [sloped,below]{} (root-1-0-0-0);
	\draw[->] (root-1-0-0) -- node [sloped]{} (root-1-0-0-1);
	\draw[->] (root-1-0-1) -- node [sloped,below]{} (root-1-0-1-0);
	\draw[->] (root-1-0-1) -- node [sloped]{} (root-1-0-1-1);
	\draw[->] (root-1-1-0) -- node [sloped,below]{} (root-1-1-0-0);
	\draw[->] (root-1-1-0) -- node [sloped]{} (root-1-1-0-1);
	\draw[->] (root-1-1-1) -- node [sloped,below]{} (root-1-1-1-0);
	\draw[->] (root-1-1-1) -- node [sloped]{} (root-1-1-1-1);
\end{tikzpicture}
    \caption{}
  \end{subfigure}
  \hfill
  \begin{subfigure}[t]{0.19\textwidth}
    \centering
    \begin{tikzpicture}[auto, scale=12,
	X1_NA/.style={circle,inner sep=1mm,minimum size=0.6cm,draw,black,very thick,fill=white,text=black},
	X2_1/.style={circle,inner sep=1mm,minimum size=0.6cm,draw,black,very thick,fill=orange,text=black},
    X2_2/.style={circle,inner sep=1mm,minimum size=0.6cm,draw,black,very thick,fill=purple,text=black},
	X3_1/.style={circle,inner sep=1mm,minimum size=0.6cm,draw,black,very thick,fill=yellow,text=black},
    X3_2/.style={circle,inner sep=1mm,minimum size=0.6cm,draw,black,very thick,fill=pink,text=black},
    X3_3/.style={circle,inner sep=1mm,minimum size=0.6cm,draw,black,very thick,fill=cyan,text=black},
	X4_1/.style={circle,inner sep=1mm,minimum size=0.6cm,draw,black,very thick,fill=blue,text=black},
    X4_2/.style={circle,inner sep=1mm,minimum size=0.6cm,draw,black,very thick,fill=green,text=black},
    X4_3/.style={circle,inner sep=1mm,minimum size=0.6cm,draw,black,very thick,fill=red,text=black},
     X4_3/.style={circle,inner sep=1mm,minimum size=0.6cm,draw,black,very thick,fill=red,text=black},
	 leaf/.style={circle,inner sep=1mm,minimum size=0.4cm,draw,very thick,black,fill=gray,text=black}]

	\node [X1_NA] (root) at (0.000000, 0.500000*\dd)	{};
	\node [X2_1] (root-0) at (0.2*\ddd, 0.233333*\dd)	{};
	\node [X2_2] (root-1) at (0.2*\ddd, 0.766667*\dd)	{};
	\node [X3_1] (root-0-0) at (0.4*\ddd, 0.100000*\dd)	{};
	\node [X3_2] (root-0-1) at (0.4*\ddd, 0.366667*\dd)	{};
	\node [X3_1] (root-1-0) at (0.4*\ddd, 0.633333*\dd)	{};
	\node [X3_2] (root-1-1) at (0.4*\ddd, 0.900000*\dd)	{};
	\node [X4_1] (root-0-0-0) at (0.6*\ddd, 0.033333*\dd)	{};
	\node [X4_1] (root-0-0-1) at (0.6*\ddd, 0.166667*\dd)	{};
	\node [X4_2] (root-0-1-0) at (0.60000*\ddd, 0.300000*\dd)	{};
	\node [X4_2] (root-0-1-1) at (0.60000*\ddd, 0.433333*\dd)	{};
	\node [X4_1] (root-1-0-0) at (0.60000*\ddd, 0.566667*\dd)	{};
	\node [X4_1] (root-1-0-1) at (0.60000*\ddd, 0.700000*\dd)	{};
	\node [X4_3] (root-1-1-0) at (0.60000*\ddd, 0.833333*\dd)	{};
	\node [X4_3] (root-1-1-1) at (0.60000*\ddd, 0.966667*\dd)	{};
	\node [leaf] (root-0-0-0-0) at (0.8000000*\ddd, 0.000000*\dd)	{};
	\node [leaf] (root-0-0-0-1) at (0.8000000*\ddd, 0.066667*\dd)	{};
	\node [leaf] (root-0-0-1-0) at (0.8000000*\ddd, 0.133333*\dd)	{};
	\node [leaf] (root-0-0-1-1) at (0.8000000*\ddd, 0.200000*\dd)	{};
	\node [leaf] (root-0-1-0-0) at (0.8000000*\ddd, 0.266667*\dd)	{};
	\node [leaf] (root-0-1-0-1) at (0.8000000*\ddd, 0.333333*\dd)	{};
	\node [leaf] (root-0-1-1-0) at (0.8000000*\ddd, 0.400000*\dd)	{};
	\node [leaf] (root-0-1-1-1) at (0.8000000*\ddd, 0.466667*\dd)	{};
	\node [leaf] (root-1-0-0-0) at (0.8000000*\ddd, 0.533333*\dd)	{};
	\node [leaf] (root-1-0-0-1) at (0.8000000*\ddd, 0.600000*\dd)	{};
	\node [leaf] (root-1-0-1-0) at (0.8000000*\ddd, 0.666667*\dd)	{};
	\node [leaf] (root-1-0-1-1) at (0.8000000*\ddd, 0.733333*\dd)	{};
	\node [leaf] (root-1-1-0-0) at (0.8000000*\ddd, 0.800000*\dd)	{};
	\node [leaf] (root-1-1-0-1) at (0.8000000*\ddd, 0.866667*\dd)	{};
	\node [leaf] (root-1-1-1-0) at (0.8000000*\ddd, 0.933333*\dd)	{};
	\node [leaf] (root-1-1-1-1) at (0.8000000*\ddd, 1.000000*\dd)	{};

	\draw[->] (root) -- node [sloped,below]{} (root-0);
	\draw[->] (root) -- node [sloped]{} (root-1);
	\draw[->] (root-0) -- node [sloped,below]{} (root-0-0);
	\draw[->] (root-0) -- node [sloped]{} (root-0-1);
	\draw[->] (root-1) -- node [sloped,below]{} (root-1-0);
	\draw[->] (root-1) -- node [sloped]{} (root-1-1);
	\draw[->] (root-0-0) -- node [sloped,below]{} (root-0-0-0);
	\draw[->] (root-0-0) -- node [sloped]{} (root-0-0-1);
	\draw[->] (root-0-1) -- node [sloped,below]{} (root-0-1-0);
	\draw[->] (root-0-1) -- node [sloped]{} (root-0-1-1);
	\draw[->] (root-1-0) -- node [sloped,below]{} (root-1-0-0);
	\draw[->] (root-1-0) -- node [sloped]{} (root-1-0-1);
	\draw[->] (root-1-1) -- node [sloped,below]{} (root-1-1-0);
	\draw[->] (root-1-1) -- node [sloped]{} (root-1-1-1);
	\draw[->] (root-0-0-0) -- node [sloped,below]{} (root-0-0-0-0);
	\draw[->] (root-0-0-0) -- node [sloped]{} (root-0-0-0-1);
	\draw[->] (root-0-0-1) -- node [sloped,below]{} (root-0-0-1-0);
	\draw[->] (root-0-0-1) -- node [sloped]{} (root-0-0-1-1);
	\draw[->] (root-0-1-0) -- node [sloped,below]{} (root-0-1-0-0);
	\draw[->] (root-0-1-0) -- node [sloped]{} (root-0-1-0-1);
	\draw[->] (root-0-1-1) -- node [sloped,below]{} (root-0-1-1-0);
	\draw[->] (root-0-1-1) -- node [sloped]{} (root-0-1-1-1);
	\draw[->] (root-1-0-0) -- node [sloped,below]{} (root-1-0-0-0);
	\draw[->] (root-1-0-0) -- node [sloped]{} (root-1-0-0-1);
	\draw[->] (root-1-0-1) -- node [sloped,below]{} (root-1-0-1-0);
	\draw[->] (root-1-0-1) -- node [sloped]{} (root-1-0-1-1);
	\draw[->] (root-1-1-0) -- node [sloped,below]{} (root-1-1-0-0);
	\draw[->] (root-1-1-0) -- node [sloped]{} (root-1-1-0-1);
	\draw[->] (root-1-1-1) -- node [sloped,below]{} (root-1-1-1-0);
	\draw[->] (root-1-1-1) -- node [sloped]{} (root-1-1-1-1);
\end{tikzpicture}
    \caption{}
  \end{subfigure}
\caption{Simulated data. (a) true staged tree used to simulate the data; (b) histogram of simulated continuous covariates; (c) estimated staged tree (VI) under the Hamming distance prior; (d) estimated staged tree (VI) under the prior using both Hamming and continuous covariates.}
  \label{fig:Simulated_PPMx}
\end{figure}
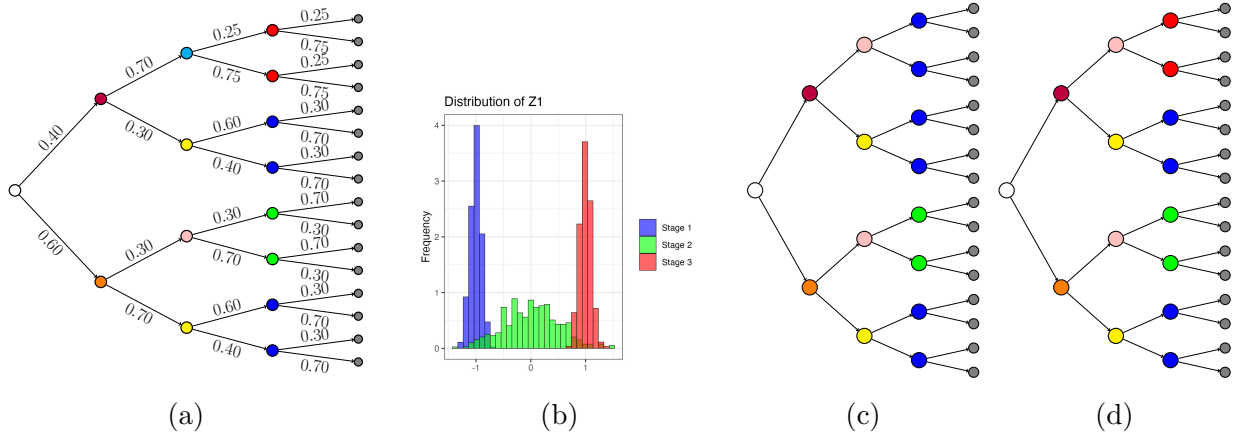

We illustrate the advantage of covariate-dependent priors with a simulated example based on a staged tree model with stage-specific probabilities. A continuous covariate is generated according to the true staging of variable $X_3$, with each stage associated with a normal distribution having distinct mean and variance. Figures~\ref{fig:Simulated_PPMx}(a)–(b) display the data-generating staged tree and the resulting covariate distribution. We generate $n = 500$ observations and estimate the model using the MCMC algorithm described in Supplementary Section 1, 
running 2000 iterations with a burn-in of 1000 and no thinning, yielding 1000 posterior samples. We compare two variants: one using only the Hamming distance ($\kappa = 1$, $\xi = 0.25$), and one incorporating the continuous covariate via a covariate-dependent prior ($\lambda_z = 0.5$). Figures~\ref{fig:Simulated_PPMx}(c)–(d) show the estimated staged trees under each model. The covariate-informed approach accurately recovers the true stage structure, while the Hamming-only model fails to separate contexts that have similar but distinct conditional distributions, which were purposely designed to be challenging to distinguish based on tree position alone.

We evaluate the impact of the hyperparameter \(\lambda_Z\), which regulates the influence of continuous covariates in the prior, through a targeted sensitivity analysis. Using the staged tree shown in Supplementary Figure S1, 
we simulate 1000 observations. Two continuous covariates are then generated to be informative about the stage assignments of $X_3$ and $X_4$ (stage-specific Gaussian means with a common variance), so that $Z_1$ aligns with stages of $X_3$ and $Z_2$ with stages of $X_4$. The MCMC algorithm is run for 2000 iterations, discarding the first 1000 as burn-in. We explore \(\lambda_{Z_1}, \lambda_{Z_2} \in \{0, 0.25, 0.5, 1, 2.5, 5\}\), yielding 36 distinct scenarios. To assess model performance, we compute the \emph{normalized Hamming distance} (the proportion of stage assignments that must be changed to recover the true model) and the \emph{Rand index} for each variable with learned stages (\(X_3\), \(X_4\), \(X_5\)). The results, shown in Supplementary Figure S3, 
confirm that larger values of \(\lambda_{Z_1}\) improve recovery for \(X_3\), while larger \(\lambda_{Z_2}\) improve estimation for \(X_4\). Performance for \(X_5\) remains consistently high across all settings, indicating robustness to irrelevant covariate information.

\section{Posterior Inference}
\label{sec:posterio_inference}

We develop a MCMC algorithm for posterior inference for staged trees equipped with the novel PPMx priors. From Equations~\eqref{eq:marginal} and~\eqref{eq:priorfact}, it follows that
\begin{equation}
\label{eq:posti}
 \log P(\mathcal{T}_{\bm{X}}^{\bm{\rho}_C}|\mathcal{D}) = \sum_{i\in C}  \log P(\mathcal{T}_{\bm{X}}^{\rho_i}|\mathcal{D}),
\end{equation}
and hence the partitions at different depths of the tree can be estimated independently. We therefore fix \( i \in C \) and focus on the posterior \( P(\mathcal{T}_{\bm{X}}^{\rho_i}|\mathcal{D}) \). Although the decomposability of the posterior distribution under structure modularity has long been recognized~\citep{freeman2011bayesian}, the only estimation approach available to date is a greedy agglomerative algorithm that returns a MAP estimate. A recent exception is the method proposed in \citet{shenvi2024beyond}, which samples partitions with a fixed number of blocks using Stan and computes posterior probabilities of stage membership. However, final estimates are obtained via hard allocation, thereby discarding the uncertainty captured in the posterior distribution.

Recall that since the marginal likelihood where the parameters $\bm\theta$ are integrated out is available in closed-form in Equation \eqref{eq:marginal1}, we can directly and uniquely sample the partitions $\rho_i$ via a collapsed sampler. Our MCMC algorithm consists of two  move types for each iteration: first, we employ the sampling scheme of \citet{neal2000markov} (algorithm 2), based on the popular P\'{o}lya urn scheme; second, we employ a split-and-merge move. The details are given in Supplementary Section 1. 

\subsection{Posterior Summaries}
The output of the MCMC algorithm is a collection of stage membership indicators approximately drawn from the posterior distribution in Equation~\eqref{eq:posti}. From these samples, we derive posterior summaries of the stage structure, quantify associated uncertainty, and estimate ATEs.

\subsubsection{The Staged Tree Estimate}

Consider a posterior sample of size $R$. We obtain partitions $\rho_i^{(1)}, \dots, \rho_i^{(R)}$ of $\mathbb{X}_{[i-1]}$, each specified by stage membership indicators $g_{\bm{x}_{[i-1]}}^{(r)}$, for $r = 1, \dots, R$. A standard approach to summarizing this output is to construct a posterior dissimilarity matrix $D$, where each entry $(\bm{x}_{[i-1]}, \bm{x}_{[i-1]}')$ represents the estimated posterior probability that the two contexts belong to different stages:
\[
\hat{P}\left(g_{\bm{x}_{[i-1]}}\neq g_{\bm{x}_{[i-1]}'}|\mathcal{D}\right)=\frac{1}{R}\sum_{r=1}^R\mathbbm{1}(g_{\bm{x}_{[i-1]}}^{(r)}\neq g_{\bm{x}_{[i-1]}'}^{(r)})
\]
A point estimate $\hat{\rho}^i$ can then be obtained by clustering together $\bm{x}_{[i-1]}$ and $\bm{x}_{[i-1]}'$ whenever their dissimilarity falls below a fixed threshold, such as 0.5~\citep{leonelli2024robust}. However, this rule is highly sensitive to the choice of threshold, and no principled guidance exists for its selection, which may compromise the robustness and interpretability of the resulting model. Instead, we adopt the approach of \citet{wade2018bayesian}, which selects a point estimate $\hat{\rho}_i$ by minimizing the posterior expectation of a loss function $L$ comparing candidate partitions to the unknown true partition:
\begin{equation}
\label{eq:opt}
\hat{\rho}_i=\arg\min_{\hat{\rho}_i}\mathbb{E}(L(\rho_i,\hat{\rho}_i)|\mathcal{D})\approx\arg\min_{\hat{\rho}_i}\frac{1}{R}\sum_{r=1}^RL(\rho_i^{(r)},\hat{\rho}_i)
\end{equation}
Popular choices for the loss function include Binder's loss~\citep{binder1978bayesian} and variation of information~\citep{meilua2007comparing}. In our work, we prefer the latter, as it often yields more parsimonious staged trees. The minimization in Equation~\eqref{eq:opt} is computationally challenging, and we employ the SALSO algorithm~\citep{dahl2022search} for its efficient solution.

Understanding the uncertainty associated with the estimated staged tree is essential for assessing the robustness of the inferred structure and the credibility of context-specific independence statements. While some methods for staged trees rely on model averaging~\citep{strong2022bayesian} or bootstrap-based summaries~\citep{leonelli2024robust}, our fully Bayesian framework naturally provides uncertainty quantification through the posterior sample. One intuitive approach is to visualize the posterior dissimilarity matrix. To move beyond qualitative inspection, we follow \citet{wade2018bayesian} and summarize uncertainty with a \emph{credible ball} around the point estimate, defined as the smallest set of partitions (under a chosen loss) containing a fixed proportion of posterior mass. This compactly highlights high-probability alternatives and clarifies which context-specific independencies are stable versus variable across plausible models.

\subsubsection{Estimating Causal Effects}

To estimate causal effects from the posterior output of our collapsed sampler, we first recover the stage-specific multinomial parameters $\bm{\theta}$, which are integrated out during inference. For each sampled partition $\bm{\rho}_C^{(r)}$, we compute the posterior mean of the stage probabilities using standard conjugate updating under a Dirichlet–Multinomial model. We then construct the corresponding causal staged tree by intervening on the treatment variable, as described in Section~\ref{sec:setup}. From each posterior sample, we compute both the average treatment effect (ATE) using the standardization formula in Equation~\eqref{eq:standardization}, and the conditional average treatment effects (CATEs) for each covariate profile by evaluating the difference in outcome probabilities under treatment and control. This yields posterior samples $\text{ATE}^{(1)}, \dots, \text{ATE}^{(R)}$ and $\text{CATE}_{\bm{z}}^{(1)}, \dots, \text{CATE}_{\bm{z}}^{(R)}$ for all observed covariate configurations $\bm{z}$. These posterior draws allow for full uncertainty quantification through summary statistics such as means, credible intervals, and tail probabilities. This approach ensures that both parameter and structural uncertainty are fully propagated into causal effect estimation, avoiding the bias and overconfidence typical of post-selection inference.

\begin{figure}
    \centering
\includegraphics[width=0.75\linewidth]{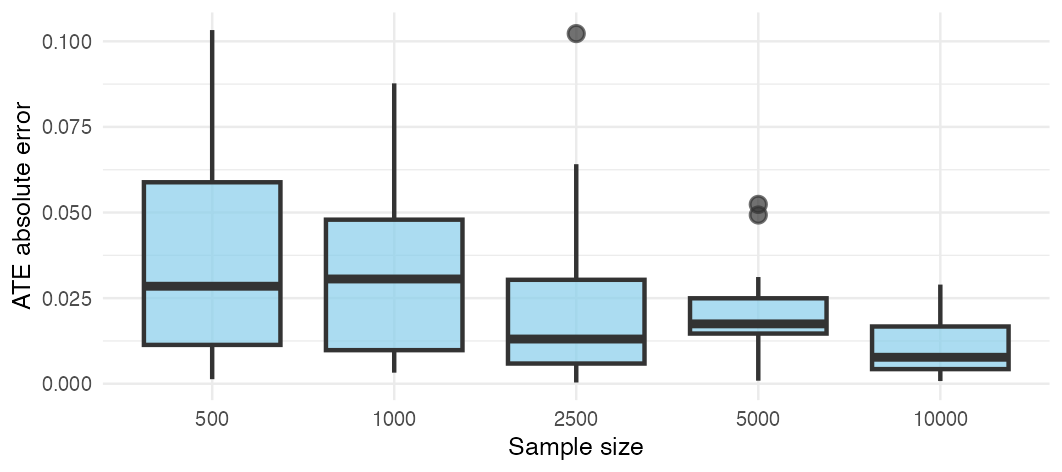}
 \caption{Effect of sample size in the consistency simulation study. Absolute ATE error across $n$; summaries over 25 replicates per $n$.}
    \label{fig:ate_sim}
\end{figure}

We demonstrate the consistency of our methodology in a small simulation study. We generate data from the staged tree in Supplementary Section 3, 
where $X_4$ is taken as the treatment and $X_5$ as the outcome. The true ATE implied by this data-generating process is $-0.1246$, while the CATEs vary across the 16 covariate profiles, taking positive, negative, or null values depending on the specific configuration. For each sample size $n \in \{500,1000,2500,5000,10000\}$ we generate 25 independent datasets, estimate the staged tree structure with our collapsed sampler, and compute both the ATE and CATEs. Figure~\ref{fig:ate_sim} summarizes the distribution of the absolute error of the ATE across sample sizes, showing a clear contraction of both the median error and its variability as $n$ increases. This confirms that our procedure reliably recovers the true causal effect, with errors quickly shrinking toward zero. Full results for the CATEs, reported in Supplementary Section 3, 
reinforce this conclusion: while some profiles converge more slowly, reflecting their lower frequency in the sample, the overall pattern is the same, with estimation errors decreasing systematically as the sample size grows. These results highlight both the consistency of our Bayesian staged tree methodology and its ability to propagate structural uncertainty when estimating heterogeneous causal effects.

\begin{table}
\centering
\caption{Median absolute ATE error with standard deviation (in parentheses) across sample sizes $n \in \{500,1000,5000\}$ and imbalance levels $q \in \{0.0,0.5,0.8\}$ under the two probability-generation schemes (\textit{exp} and \textit{unif}).  \textbf{Bold} marks the lowest median, while \textit{italics} indicate the second–lowest.\label{table:ate_comparison}}
\scalebox{0.56}{%
\begin{tabular}{c c r c c c c c c}
\toprule
\multicolumn{3}{c}{} & \multicolumn{6}{c}{Methods: median (sd)} \\
\cmidrule(lr){4-9} Gen & $q$ & $N$ & Tree Bayes & BHC & CS-BHC & DAG Tabu & DAG PC & DAG Bayes \\
\midrule

  \multirow{9}{*}{exp} & \multirow{3}{*}{0.0} &  500 & 0.041 (0.038) & \textbf{0.032} (0.040) & 0.039 (0.039) & 0.039 (0.035) & 0.039 (0.040) & \textit{0.037} (0.037) \\
                        &                       & 1000 & \textit{0.016} (0.021) & 0.020 (0.025) & 0.027 (0.026) & \textbf{0.014} (0.027) & 0.022 (0.030) & \textit{0.016} (0.031) \\
                        &                       & 5000 & \textit{0.005} (0.011) & 0.008 (0.010) & \textbf{0.003} (0.013) & 0.007 (0.010) & 0.008 (0.031) & 0.007 (0.010) \\
\cmidrule(lr){2-9}
                        & \multirow{3}{*}{0.5} &  500 & \textit{0.059} (0.138) & \textbf{0.055} (0.134) & 0.072 (0.131) & 0.065 (0.138) & \textit{0.059} (0.137) & 0.065 (0.139) \\
                        &                       & 1000 & \textbf{0.041} (0.115) & 0.070 (0.101) & 0.068 (0.104) & \textit{0.052} (0.114) & 0.058 (0.113) & \textit{0.052} (0.122) \\
                        &                       & 5000 & \textbf{0.021} (0.156) & 0.038 (0.152) & \textit{0.024} (0.159) & \textit{0.024} (0.156) & 0.026 (0.158) & \textit{0.024} (0.154) \\
\cmidrule(lr){2-9}
                        & \multirow{3}{*}{0.8} &  500 & \textbf{0.033} (0.114) & 0.061 (0.111) & 0.063 (0.112) & 0.051 (0.111) & 0.049 (0.114) & \textit{0.047} (0.111) \\
                        &                       & 1000 & \textbf{0.039} (0.192) & 0.063 (0.197) & 0.057 (0.194) & \textit{0.054} (0.186) & \textit{0.054} (0.186) & \textit{0.054} (0.186) \\
                        &                       & 5000 & 0.036 (0.130) & \textbf{0.022} (0.128) & 0.027 (0.130) & 0.027 (0.130) & \textit{0.024} (0.132) & 0.027 (0.129) \\
\cmidrule(lr){1-9}
   \multirow{9}{*}{unif} & \multirow{3}{*}{0.0} &  500 & \textit{0.024} (0.033) & 0.037 (0.043) & \textbf{0.023} (0.052) & 0.027 (0.041) & 0.039 (0.039) & 0.032 (0.037) \\
                         &                       & 1000 & \textit{0.018} (0.020) & 0.026 (0.028) & 0.042 (0.033) & 0.026 (0.024) & 0.025 (0.024) & \textbf{0.015} (0.023) \\
                         &                       & 5000 & \textbf{0.007} (0.008) & 0.012 (0.012) & 0.012 (0.008) & 0.009 (0.007) & \textit{0.008} (0.015) & 0.009 (0.007) \\
\cmidrule(lr){2-9}
                         & \multirow{3}{*}{0.5} &  500 & \textbf{0.049} (0.084) & 0.062 (0.082) & 0.073 (0.091) & \textit{0.052} (0.093) & \textit{0.052} (0.094) & \textit{0.052} (0.093) \\
                         &                       & 1000 & \textbf{0.027} (0.131) & 0.032 (0.137) & 0.044 (0.127) & 0.034 (0.135) & 0.038 (0.134) & \textbf{0.027} (0.131) \\
                         &                       & 5000 & 0.041 (0.061) & \textbf{0.036} (0.060) & 0.038 (0.060) & \textit{0.037} (0.059) & \textit{0.037} (0.059) & \textit{0.037} (0.059) \\
\cmidrule(lr){2-9}
                         & \multirow{3}{*}{0.8} &  500 & \textbf{0.031} (0.061) & 0.044 (0.056) & \textit{0.041} (0.050) & 0.052 (0.056) & 0.052 (0.056) & 0.053 (0.056) \\
                         &                       & 1000 & \textbf{0.013} (0.037) & 0.029 (0.038) & 0.028 (0.039) & \textit{0.021} (0.042) & \textit{0.021} (0.042) & \textit{0.021} (0.042) \\
                         &                       & 5000 & \textit{0.027} (0.051) & 0.029 (0.051) & \textbf{0.026} (0.051) & 0.043 (0.051) & 0.043 (0.051) & 0.043 (0.051) \\
\bottomrule
\end{tabular}%
}
\end{table}

In addition, we conduct a comparative study against established competitors in the graphical models space. Specifically, we generate data from random staged trees over six binary variables. For each variable, a parent set is selected uniformly at random, and the corresponding staged tree coloring is obtained. Stages are then merged at random with probability $q \in \{0,0.5,0.8\}$, and stage probabilities are sampled either from normalized exponential draws (yielding probabilities uniformly distributed on the simplex) or from normalized uniform draws. From each staged tree, we sample datasets of sizes $n \in \{500,1000,2500\}$, repeating the procedure $25$ times per configuration. We compare six estimators of the ATE based on graphical models by evaluating their \emph{absolute ATE error}. Three are staged tree–based: our Bayesian algorithm (with default hyperparameters and no covariates), the BHC approach of \citet{Carli2020}, and the CS-BHC algorithm of \citet{varando2024staged} that only searches for context-specific independences. The remaining three are DAG–based: score-based structure learning with tabu search, constraint-based estimation via the PC algorithm, and the Bayesian partition algorithm of \texttt{BiDAG} \citep{suter2023bayesian}, from which we obtain posterior mean ATE estimates. The results, reported in Table~\ref{table:ate_comparison}, show that our Bayesian staged tree approach achieves the lowest or second-lowest absolute ATE error in all but three scenarios, and is the best performer in 50\% of cases. Notably, while non-Bayesian staged tree methods are often outperformed by the Bayesian procedure proposed in this study, they have themselves been shown to be competitive with standard causal effect estimation techniques \citep{varando2025causal}. This reinforces the conclusion that our Bayesian approach combines the interpretability of staged trees with improved accuracy, offering a strong contribution to causal effect estimation.

\section{Applications of Staged Tree Causal Inference}\label{sec:rhc}

We present in this section two real-world applications of staged tree causal inference. Specifically, we first study a dataset on the effect of electronic fetal monitoring on cesarean section rates in Section \ref{sec:efm}, and then investigate the effect of anthracycline treatment on cardiac dysfunction in breast cancer patients in Section \ref{sec:bc}.

\subsection{Cesarean Section Data}\label{sec:efm}

We implement the proposed approach using data from an observational benchmark in causal inference, studying the effect of electronic fetal monitoring on the likelihood of cesarean section. The dataset, originally collected by \citet{neutra1980effect} and reformatted by \citet{richardson2017modeling}, contains observations on \(14{,}484\) deliveries recorded at Beth Israel Hospital in Boston between 1970 and 1975. All variables are binary and coded as \texttt{y}/\texttt{n}. For ease of exposition we exclude the year of delivery. The outcome variable of interest is \texttt{cesarean} (C), the treatment is \texttt{monitor} (M), and the covariates include \texttt{nullipar} (N, indicating nulliparity), \texttt{breech} (B, indicating malpresentation), and \texttt{arrest} (A, indicating arrest of labor progression). These three variables are known confounders of the relationship between treatment and cesarean outcomes, and have been consistently included in prior analyses. We learn a staged event tree over the variable ordering \texttt{nullipar}, \texttt{breech}, \texttt{arrest}, \texttt{monitor}, \texttt{cesarean}, reflecting the assumed temporal structure of the delivery process. We focus on learning the stage structure of the treatment (\texttt{monitor}) and outcome (\texttt{cesarean}) variables, initializing all vertices in separate stages without any prior grouping. Note that no continuous covariates are available in this study, and therefore we employ the version of the model described in Eq.~\eqref{eq:prior}. We run the MCMC algorithm for 10000 iterations, after a burn-in of 1000, and collect 2000 samples for posterior inference, thinning every 5th iteration.

\begin{figure}
    \centering
    \begin{subfigure}{0.49\linewidth}
        \centering
        \includegraphics[width=\linewidth]{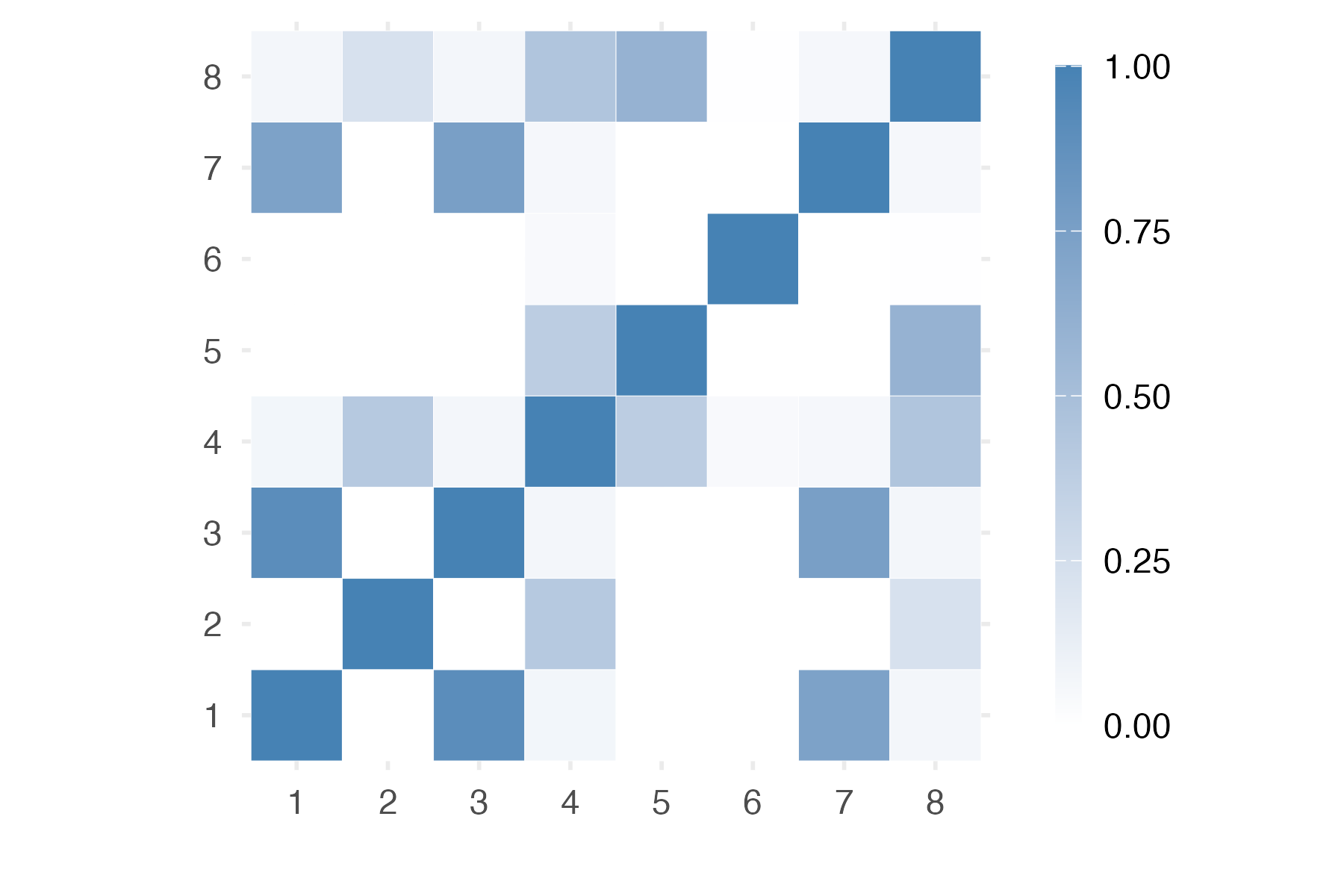}
        \caption{Monitor}
        \label{fig:sim1}
    \end{subfigure}    \begin{subfigure}{0.49\linewidth}
        \centering
        \includegraphics[width=\linewidth]{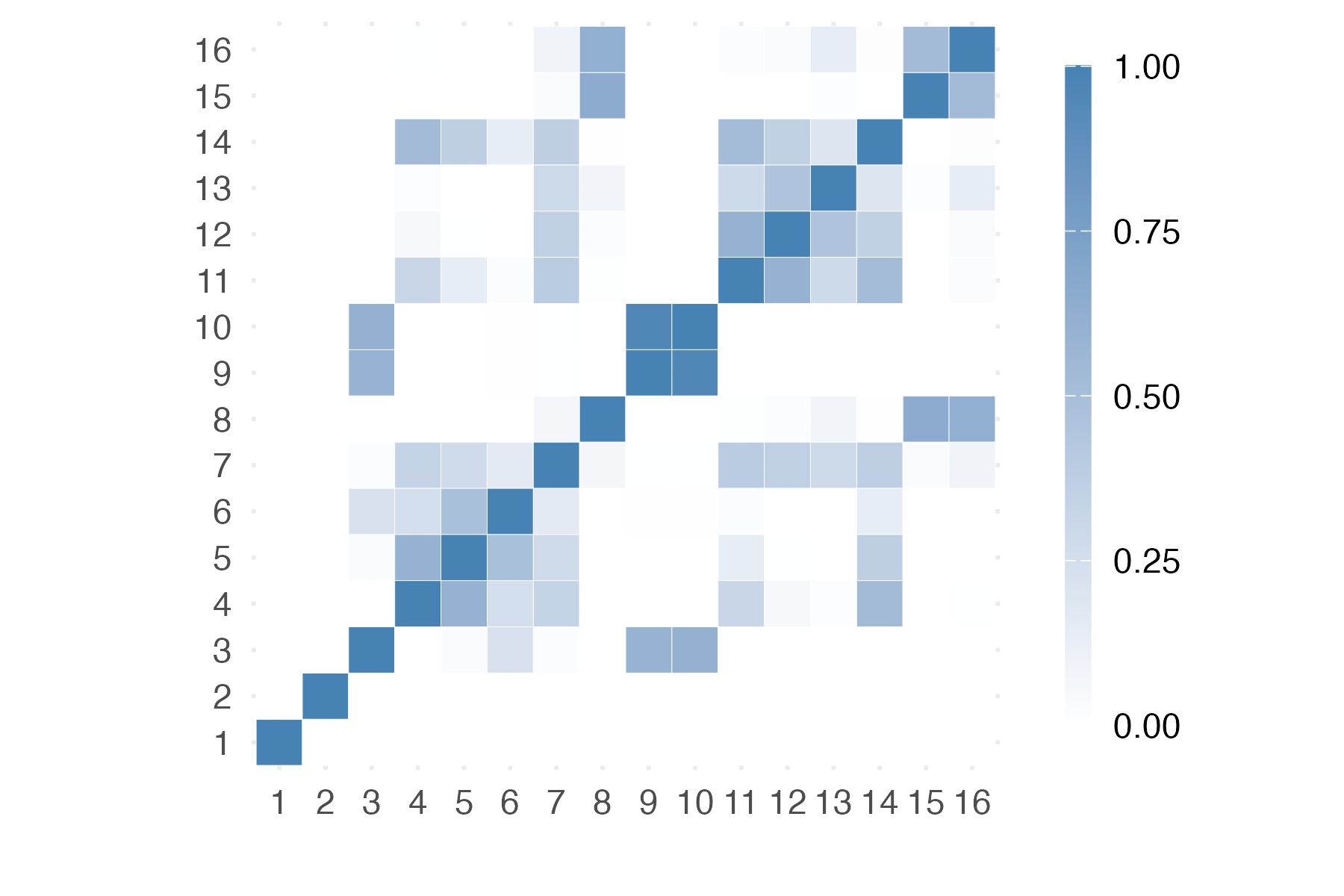}
        \caption{Cesarean}
        \label{fig:sim2}
    \end{subfigure}
    
   \caption{Cesarean Section data. Posterior co-clustering probabilities for the variables \textit{monitor} and \textit{cesarean}. Each heatmap shows the posterior probability that any two vertices belong to the same stage, based on MCMC samples. Vertices are numbered from bottom to top of the staged tree.}

    \label{fig:similarity}
\end{figure}

Figure~\ref{fig:similarity} shows the posterior co-clustering probabilities for the vertices of \textit{monitor} and \textit{cesarean}. The matrices indicate a relatively sparse structure, with most vertices rarely grouped together across posterior samples. Nonetheless, a few blocks of consistently high similarity emerge, revealing subsets of vertices that are repeatedly clustered together, suggesting strong evidence for shared behavior in those subgroups.

\begin{figure}
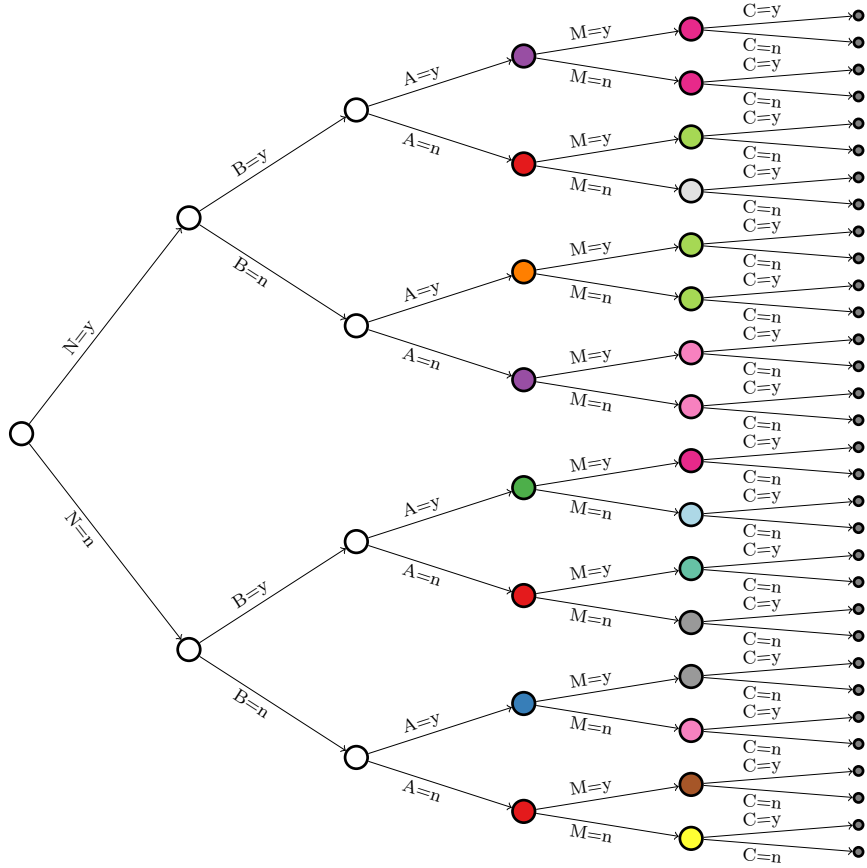

    \centering
\includestandalone[width=0.7\linewidth]{placeholder}
\caption{Cesarean Section data. Posterior staged tree point estimate for the cesarean data, obtained by minimizing the expected variation of information loss over the MCMC output. Edge labels indicate the realized outcomes of the corresponding variables.}
    \label{fig:placeholder}
\end{figure}

The staged tree shown in Figure~\ref{fig:placeholder}, obtained by minimizing the expected variation of information across posterior samples, highlights several context-specific independence structures. For the treatment variable \texttt{monitor}, women with \texttt{arrest} = n are generally grouped in the same stage (red vertices), except for those with \texttt{nullipar} = y and \texttt{breech} = n. This suggests the context-specific independencies \( M \perp\!\!\!\perp N \mid A = \text{n}, B = \text{y} \) and \( M \perp\!\!\!\perp B \mid A = \text{n}, N = \text{n} \). For the outcome \textit{cesarean}, the estimated structure indicates that treatment has little effect for most women with \texttt{nullipar} = y (e.g., pink, light green, and magenta vertices), except when \( B = \text{y}, A = \text{n} \), revealing heterogeneity in treatment response. These patterns demonstrate the ability of staged tree models to detect nuanced forms of dependence and conditional independence.

Further insight into the uncertainty of the learned staged tree structure is provided by the $95\%$ credible balls, summarized in Supplementary Section 3. 
For simplicity, we focus our interpretation on the variable \texttt{monitor}, though analogous conclusions apply to \texttt{cesarean}. The vertical lower bound of the credible ball contains three partitions, each consisting of seven stages. Since the vertical lower bound collects the most complex stage configurations among those within the credible region, its structure allows us to reject the hypothesis of full dependence of treatment assignment on all covariates. Conversely, the vertical upper bound comprises a single partition with only four stages (1, 2, 1, 1, 3, 4, 1, 1), representing the simplest admissible structure in the credible region. Its configuration permits us to reject a wide range of simplified models that would imply any symmetric conditional independencies involving \texttt{monitor} and the covariates. Overall, the structure of the credible ball suggests that the relationship between \texttt{monitor} and the covariates cannot be adequately captured by a standard DAG. However, some context-specific statements remain compatible with this structure. For example, the third, fourth, seventh, and eighth vertices (counting from the bottom of the tree) always share the same stage, indicating that the context-specific independence \( M \perp\!\!\!\perp (A, N) \mid B = \text{y} \) cannot be ruled out.

\begin{table}
\caption{Cesarean Section data. Posterior summaries of the CATE for the cesarean data across each covariate profile: posterior mean, standard deviation, and probability that the effect is positive, null, or negative.}
\label{tab:cate_summaries}
\centering
\scalebox{0.7}{
\begin{tabular}{lrrrrr}
\toprule
\textbf{Covariate Profile} & \textbf{Mean(CATE)} & \textbf{SD(CATE)} & \textbf{P(CATE $>$ 0)} & \textbf{P(CATE = 0)} & \textbf{P(CATE $<$ 0)} \\
\midrule
N=n, B=n, A=n &  $0.0156$ & $0.0001$ & $1.0000 $& $0.0000$ & $0.0000$ \\
N=n, B=n, A=y &  $0.1991$ & $0.0408$ &$ 0.9990$ & $0.0010$ & $0.0000$ \\
N=n, B=y, A=n & $-0.0579$ & $0.0643$ &$0.0165$ & $0.4910$ & $0.4925$ \\
N=n, B=y, A=y &  $0.3542$ & $0.1376$ & $0.9285$ & $0.0655$ & $0.0060$ \\
N=y, B=n, A=n &  $0.0003$ & $0.0015$ & $0.0460$ & $0.9540$ & $0.0000$ \\
N=y, B=n, A=y &  $0.0298$ & $0.0395$ & $0.3930$ & $0.5990$ & $0.0080$ \\
N=y, B=y, A=n & $-0.1156$ & $0.0724$ & $0.0000$ & $0.1935$&$ 0.8065$ \\
N=y, B=y, A=y & $-0.0969$ & $0.1209$ & $0.0045$ & $0.5250$ & $0.4705$ \\
\bottomrule
\end{tabular}}
\end{table}

We now turn to the estimation of causal effects. The posterior distribution of the ATE reveals a consistently positive effect of electronic fetal monitoring on cesarean section rates, with a posterior mean of $0.0127$, standard deviation of $0.0041$, and a $95\%$ credible interval of $(0.0060,\ 0.0208)$. This result is in line with previous findings based on parametric modeling~\citep{richardson2017modeling} and confirms that, on average, use of monitoring is associated with a higher probability of cesarean delivery. However, the posterior distributions of the CATEs, summarized in Table~\ref{tab:cate_summaries}, highlight strong heterogeneity across subgroups. For instance, the effect of EFM is close to zero or negative for all women with \texttt{nullipar} = y, while it is positive and significant for those with \texttt{nullipar} = n and \texttt{arrest} = y. These patterns echo prior domain knowledge~\citep{neutra1980effect} and mirror the subgroup-specific findings in~\cite{richardson2017modeling}, despite our model relying only on categorical covariates and not accounting for time-varying information. The staged tree model enables posterior inference of causal effects while maintaining transparency in the structure of dependencies, which facilitates interpretation and communication of results.

\subsection{Breast Cancer Data}\label{sec:bc}

We now consider a second real-world application involving the risk of cardiac dysfunction following oncologic treatment in women with breast cancer. The dataset, originally introduced by \citet{pineiro2023cardiotoxicity}, consists of clinical and imaging variables for $531$ patients diagnosed with HER2+ breast cancer and treated at the University Hospital of A Coruña between 2007 and 2021. Of these, $54$ women (approximately $10\%$) developed cancer therapy-related cardiac dysfunction (CTRCD) during follow-up. For our analysis, we focus on a subset of $474$ patients with complete observations for the selected variables.

Our goal is to estimate the causal effect of anthracycline-based therapy (\texttt{AC}) on the risk of developing CTRCD. To mitigate positivity violations due to sample sparsity, we restrict our discrete covariates to three binary variables with sufficiently balanced distributions: hypertension (\texttt{HTA}), dyslipidemia (\texttt{DL}), and past treatment history (\texttt{PT}). The latter is constructed by aggregating prior exposure to antiHER2 therapy, anthracyclines, and radiotherapy. These covariates are selected for their clinical relevance, as discussed in \citet{castelletti2024bayesian}, and to ensure empirical support across all strata. To further account for individual heterogeneity, we incorporate four continuous covariates, age, body mass index (BMI), heart rate, and baseline left ventricular ejection fraction (LVEF), also identified as key predictors in \citet{castelletti2024bayesian}. All continuous covariates are standardized prior to analysis. We adopt the PPMx-based framework described in Section~\ref{sec:model}, using a product partition prior over the vertices of the treatment and outcome variables. The staged tree is learned over the variable ordering \texttt{PT}, \texttt{HTA}, \texttt{DL}, \texttt{AC}, \texttt{CTRCD}, with continuous covariates guiding the clustering through the distance-based component of the prior. We collect $2000$ posterior samples, thinning every $5$ iterations after a burn-in of $1000$, and fix the covariate-weight hyperparameter $\lambda_z = 1$ for all continuous covariates.

\begin{figure}
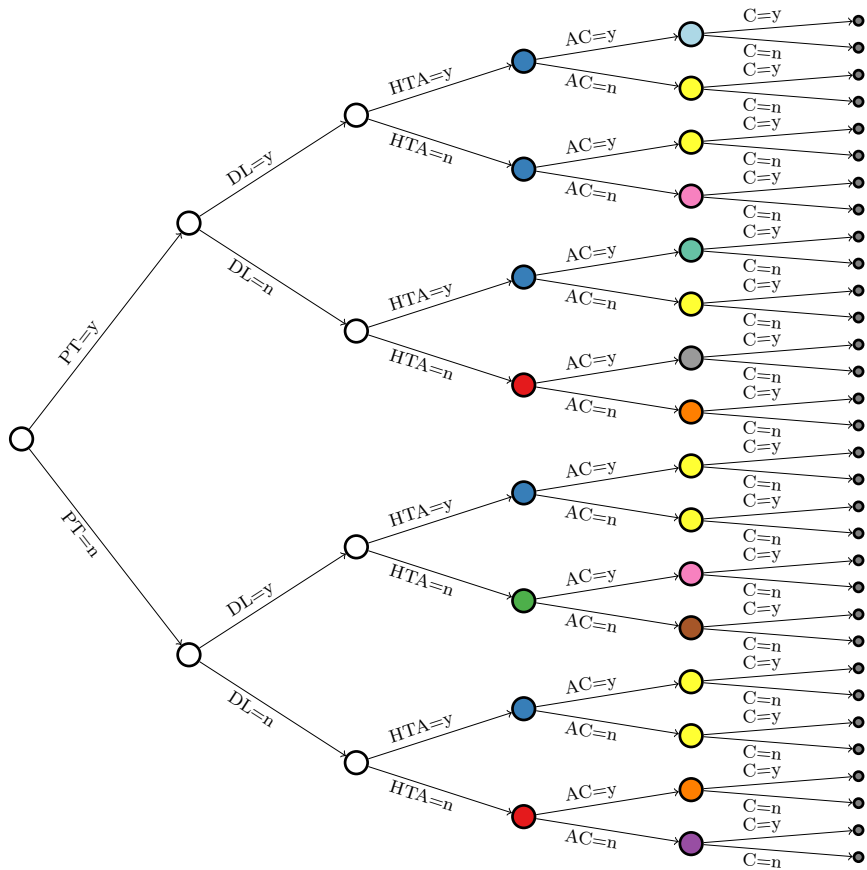

    \centering
    \includestandalone[width=0.7\linewidth]{cttree}
    \caption{Breast Cancer data. Posterior staged tree point estimate for the CTRCD (shortened to C) data, obtained by minimizing the expected variation of information loss over the MCMC output. Edge labels indicate the realized outcomes of the corresponding variables.
}
    \label{fig:cttree}
\end{figure}

\begin{table}
\caption{Breast Cancer data. Summary of continuous covariates (mean and standard deviation), sample sizes, and probability of outcome = Yes, for each stage of AC and CTRCD. Stage numbers are annotated with their corresponding stage colors.}
\label{tab:covariate_summary}
\centering
\scalebox{0.78}{
\begin{tabular}{llrrrrrrrrrr}
\toprule
\multirow{2}{*}{Variable} & \multirow{2}{*}{Stage (Color)} 
& \multicolumn{2}{c}{Age} & \multicolumn{2}{c}{BMI} & \multicolumn{2}{c}{Heart Rate} & \multicolumn{2}{c}{LVEF} & \multirow{2}{*}{$n$} & \multirow{2}{*}{$P(\text{Yes})$} \\
\cmidrule(lr){3-4} \cmidrule(lr){5-6} \cmidrule(lr){7-8} \cmidrule(lr){9-10}
 &  & Mean & SD & Mean & SD & Mean & SD & Mean & SD & & \\
\midrule
\multirow{3}{*}{AC}
 & 1 (red)         & 50.4 & 10.4 & 26.3 & 5.36 & 74.4 & 13.2 & 65.4 & 6.94 & 324 & 0.710 \\
 & 2 (blue)        & 64.1 &  9.46 & 29.3 & 5.47 & 73.6 & 11.6 & 66.0 & 6.48 & 112 & 0.625 \\
 & 3 (green)       & 58.3 &  9.14 & 26.2 & 4.18 & 71.9 & 12.5 & 66.0 & 6.94 &  38 & 0.683 \\
\midrule
\multirow{8}{*}{CTRCD}
 & 1 (purple)      & 52.8 & 11.8 & 27.1 & 5.59 & 71.5 & 10.2 & 65.6 & 6.71 &  68 & 0.016 \\
 & 2 (orange)      & 49.4 &  9.93 & 26.2 & 5.43 & 74.9 & 14.0 & 65.8 & 6.66 & 227 & 0.084 \\
 & 3 (yellow)      & 64.4 &  9.73 & 29.9 & 5.44 & 72.9 & 10.8 & 66.3 & 6.53 &  98 & 0.113 \\
 & 4 (brown)       & 57.2 &  8.54 & 25.2 & 3.27 & 64.8 &  8.80 & 65.3 & 6.58 &  12 & 0.008 \\
 & 5 (pink)        & 59.2 &  8.90 & 26.0 & 4.26 & 74.7 & 12.5 & 66.0 & 6.70 &  35 & 0.173 \\
 & 6 (grey)        & 52.3 &  9.66 & 25.3 & 3.94 & 77.6 & 12.1 & 62.5 & 8.92 &  29 & 0.243 \\
 & 7 (turquoise)   & 67.0 &  8.19 & 26.4 & 5.55 & 85.7 & 13.2 & 58.9 & 4.40 &   3 & 0.656 \\
 & 8 (light blue)  & 61.5 &  6.36 & 28.3 & 4.64 & 89.5 & 31.8 & 67.2 & 10.3 &   2 & 0.045 \\
\bottomrule
\end{tabular}
}
\end{table}

Figure~\ref{fig:cttree} reports the posterior staged tree point estimate for the CTRCD data, while Table~\ref{tab:covariate_summary} summarizes the stage-specific distributions of the continuous covariates together with sample sizes and probabilities of a positive outcome. Compared to the previous application, where the analysis focused exclusively on the discrete variables, the present setting allows us to highlight how continuous covariates refine the interpretation of the staging structure and yield a more nuanced view of patient risk profiles.

For the treatment variable \texttt{AC}, the staged tree again partitions patients primarily according to past treatment status and comorbidity. Patients in Stage~1 (red), who have the lowest probability of receiving AC, correspond to those with previous treatment except for the case of no additional comorbidities. These patients are on average older and have higher BMI than the rest of the cohort, and they also display lower heart rate. By contrast, Stages~2 and~3 (blue and green), where the probability of receiving AC remains high, include younger and leaner patients, with higher average heart rate. Across all three stages, LVEF remains relatively stable, suggesting that ventricular function does not drive treatment assignment in this cohort.

For the outcome \texttt{CTRCD}, the staged tree reveals three broad strata of risk that align with distinct discrete covariate profiles. The highest-risk groups are Stages~6 and~7 (grey and turquoise), which correspond to patients with previous treatment (\texttt{PT} = y) and no diagnosis of dyslipidemia (\texttt{DL} = n). Despite this common discrete profile, the two stages diverge in their continuous covariates: Stage~6 patients are the youngest group (mean age 52 years) with relatively elevated heart rate, while Stage~7 patients are the oldest group (mean age 67 years). Both groups present the lowest LVEF values across the sample, consistent with impaired cardiac function, and their probabilities of CTRCD are markedly high (24\% and 66\%, respectively). Moderate-risk stages (2, 3, and 5; orange, yellow, pink) have probabilities between 8–17\% and are characterized by different combinations of comorbidity and treatment history. Stage~2 (PT = n, DL = n) consists of younger patients with elevated heart rate; Stage~3 (PT = n, DL = y) contains older patients with higher BMI; and Stage~5 (PT = y, DL = y) represents patients of intermediate age but again with higher heart rate. Finally, the lowest-risk stages (1, 4, and 8; purple, brown, light blue) correspond to patients with no previous treatment and favorable discrete profiles, who are of middle age, with BMI near normal, heart rate within the normal range, and preserved LVEF. In these groups the probability of CTRCD remains below 5\%.

Taken together, the tree structure shows how the discrete covariates define the main partitions between high- and low-risk groups, while the continuous covariates sharpen the clinical interpretation of each subgroup. Extremes of age and reductions in LVEF identify the most vulnerable patients (Stages~6--7), heart rate helps to separate moderate-risk from low-risk profiles, and BMI plays a more modest role.

\begin{table}
\caption{Breast Cancer data. Posterior summaries of the CATE for the CTRCD data across each covariate profile: posterior mean, standard deviation, and probabilities that the effect is positive, null, or negative.
}
\label{tab:cate_bc}
\centering
\scalebox{0.68}{
\begin{tabular}{lrrrrr}
\toprule
\textbf{Covariate Profile} & \textbf{Mean(CATE)} & \textbf{SD(CATE)} & \textbf{P(CATE $>$ 0)} & \textbf{P(CATE = 0)} & \textbf{P(CATE $<$ 0)} \\
\midrule
HTA=0, DL=0, past\_treat=0 & $0.0358$ & $0.0339$ & $0.721$ & $0.204$ & $0.075$ \\
HTA=0, DL=0, past\_treat=1 & $0.0569$ & $0.0847$ & $0.504$ & $0.364$ & $0.132$ \\
HTA=0, DL=1, past\_treat=0 & $0.0258$ & $0.0584$ & $0.560$ & $0.083$ & $0.356$ \\
HTA=0, DL=1, past\_treat=1 & $0.0294$ & $0.0706$ & $0.531$ & $0.235$ & $0.234$ \\
HTA=1, DL=0, past\_treat=0 & $0.0722$ & $0.0786$ & $0.668$ & $0.218$ & $0.114$ \\
HTA=1, DL=0, past\_treat=1 & $0.1463$ & $0.1897$ & $0.721$ & $0.210$ & $0.070$ \\
HTA=1, DL=1, past\_treat=0 & $0.0200$ & $0.0474$ & $0.433$ & $0.394$ & $0.173$ \\
HTA=1, DL=1, past\_treat=1 & $0.0095$ & $0.0790$ & $0.394$ & $0.240$ & $0.366$ \\
\bottomrule
\end{tabular}
}
\end{table}

We now turn to the estimation of the causal effect of anthracycline treatment on CTRCD.  
The posterior distribution of the ATE has a mean of $0.041$ (sd $0.025$), with a $95\%$ credible interval $[-0.004, 0.088]$. The posterior probability that the effect is positive is $0.96$, indicating strong evidence for an increased risk of CTRCD following anthracycline therapy. These results are consistent with those reported by \citet{castelletti2024bayesian}, who also found a modest but consistently positive effect of anthracyclines on cardiotoxicity.  

Conditional effects across covariate profiles (Table~\ref{tab:cate_bc}) show greater heterogeneity. While most CATEs are positive, uncertainty is substantial, reflecting the smaller sample size. Patients with hypertension but no previous treatment (HTA = 1, DL = 0, PT = 0) display the strongest mean effect ($0.072$), whereas those with accumulated comorbidities (HTA = 1, DL = 1, PT = 1) have near-zero effect with wide uncertainty. This pattern echoes the two-cluster structure identified by \citet{castelletti2024bayesian}: patients with fewer baseline risk factors show stronger treatment effects, while those with multiple comorbidities yield weaker or more uncertain estimates.

\section{Conclusions}
\label{sec:concl}

This paper has introduced the first fully Bayesian framework for staged tree learning, grounded in novel prior distributions derived from PPMs. By framing the staging problem in terms of clustering, our approach enables a principled, model-based investigation of uncertainty in the learned relationships through posterior summaries and credible balls. The formulation naturally accommodates continuous covariates via covariate-dependent priors, thus avoiding ad hoc discretization, and inference is supported by an efficient MCMC scheme based on collapsed sampling with split-and-merge moves. Together, these developments establish staged trees as a flexible and computationally tractable class of models for categorical and mixed data.  

From a causal perspective, our contribution represents one of the few attempts to jointly perform causal discovery and inference within a Bayesian framework. Unlike approaches that first learn a model and subsequently estimate causal effects, our methodology integrates both steps into a unified posterior analysis, thereby avoiding the pitfalls of post-hoc inference. The two real-world applications presented here illustrate how staged trees can reveal interpretable structure in observational data and provide context-specific insights into causal effects in medical domains.  

Beyond the current setup, we note two modifications that warrant consideration. First, while classical Bayesian nonparametric approaches often place hyperpriors on the parameters of product partition models \citep[e.g.][]{escobar1995bayesian}, we opted to fix these values rather than estimate them. Doing so would render the normalizing constant of the prior intractable, requiring the use of generic algorithms such as the exchange algorithm \citep{murray2006exchange}, which involves simulating entire staged trees at each iteration. In our experiments, this substantially increased computational cost without delivering noticeable gains in practice. Moreover, our sensitivity analyses indicated that reasonable fixed choices of these parameters provide stable results, and that their influence decreases with sample size. Second, our current incorporation of continuous covariates through covariate-dependent priors does not include explicit variable selection, meaning that irrelevant covariates may still enter the prior formulation. Although preliminary attempts in this direction proved challenging due to the additional uncertainty introduced, more systematic approaches could be explored \citep{barcella2017comparative}.   

Several further directions for research emerge from this work. The Hamming-based prior we proposed is an example of incorporating external information into the clustering process. Similar ideas could be pursued in settings where multiple staged trees are estimated across geographical locations, imposing spatial coherence through spatial PPM formulations \citep[e.g.][]{page2016spatial}. More broadly, establishing a link between staged trees and product partition models opens the door to transferring recent advances in informed and dependent PPMs \citep{paganin2020centered, page2022dependent} to staged tree learning. A particularly promising avenue is the development of priors tailored to simple staged trees \citep{leonelli2024structural}, which restrict partitions at deeper levels to depend on those at earlier depths, thereby enhancing interpretability. These directions highlight the potential of staged tree models as a versatile Bayesian tool for both methodological and applied causal research.

\bibliographystyle{Chicago}

\bibliography{bib}

\appendix 
\newpage
\section{Supplementary Material}
\subsection{MCMC Algorithm}\label{SMsec:MCMC}

We report in this section the two main steps of the tailored MCMC algorithm used for posterior inference, namely the P\'{o}lya urn scheme of \citet{neal2000markov} (algorithm 2), and a split-and-merge move.

\subsubsection{P\'{o}lya Urn Step}

We introduce stage membership indicators \( g_{\bm{x}_{[i-1]}} \in \{1, \dots, M_i\} \), where \( g_{\bm{x}_{[i-1]}} = k \) if \( \bm{x}_{[i-1]} \in S_k \). Let \( \bm{g}_{-i} \) denote the current allocation of all contexts other than \( \bm{x}_{[i-1]} \), and let \( n_k = \#\{S_k \setminus \bm{x}_{[i-1]}\} \) be the number of contexts currently assigned to stage \( k \), excluding \( \bm{x}_{[i-1]} \) itself. The full conditional for \( g_{\bm{x}_{[i-1]}} \) is then proportional to:
\begin{equation}
\label{eq:ratio}
P(g_{\bm{x}_{[i-1]}}=k | \mathcal{D}, \bm{g}_{-i}) \propto \left\{
\begin{array}{ll}
n_ke^{-\xi D_k - \sum\limits_{z=1}^{q} \lambda_z \Delta^{(z)}_k} \frac{m\left(\bm{N}_{S_k\cup\bm{x}_{[i-1]}}\right)}{m\left(\bm{N}_{S_k\setminus \bm{x}_{[i-1]}}\right)}, & \mbox{if } k\in\{1,\dots,M_i\} \\
\kappa m\left(\bm{N}_{\bm{x}_{[i-1]}}\right), & \mbox{if } k=M_i+1
\end{array}
\right.
\end{equation}
where 
\begin{align}
    D_k &= \sum_{j,l\in S_k\cup \bm{x}_{[i-1]}} d_{j,l} - \sum_{j,l\in S_k\setminus  \bm{x}_{[i-1]}} d_{j,l},\label{eq:delta1}\\
    \Delta^{(z)}_k &= \sum_{j,l\in S_k\cup \bm{x}_{[i-1]}} \delta^{(z)}_{j,l} - \sum_{j,l\in S_k\setminus  \bm{x}_{[i-1]}} \delta^{(z)}_{j,l}, \quad z = 1, \dots, q.\label{eq:delta2}
\end{align}
Notice that Equations \eqref{eq:delta1} and \eqref{eq:delta2} straightforwardly simplify to
\begin{align*}
D_k &= \sum_{j\in S_k\setminus\bm{x}_{[i-1]}}d_{\bm{x}_{[i-1]},j}, \\
\Delta^{(z)}_k &= \sum_{j\in S_k\setminus\bm{x}_{[i-1]}}\delta^{(z)}_{\bm{x}_{[i-1]},j}, \quad z = 1, \dots, q,
\end{align*}
while the ratio of marginal likelihoods in Equation \eqref{eq:ratio} can be written as

\begin{equation}
\label{eq:fact}
    \frac{m\left(\bm{N}_{S_k\cup\bm{x}_{[i-1]}}\right)}{m\left(\bm{N}_{S_k\setminus \bm{x}_{[i-1]}}\right)}=\frac{\prod_{x_i\in\mathbb{X}_i}N_{\bm{x}_{[i-1]}}^{x_i}!}{|\bm{N}_{\bm{x}_{[i-1]}}|!}.
\end{equation}
A proof of this statement can be found in Supplementary Section~\ref{SMsec:proof1}.

\subsubsection{Split-And-Merge Step}

Each iteration concludes with a split-and-merge step, where two contexts \( \bm{x}_{[i-1]}, \bm{x}_{[i-1]}' \in \mathbb{X}_{[i-1]} \) are selected at random. If \( g_{\bm{x}_{[i-1]}} \neq g_{\bm{x}_{[i-1]}'}, \) a merge move is proposed; otherwise, a split move is considered. In the merge case, let $g_{\bm{x}_{[i-1]}}=k$, $g_{\bm{x}_{[i-1]}'}=k'$ and $\rho_i^*=\rho_i\cup \{S_k\cup S_{k'}\}\setminus S_k \setminus S_{k'}$. In the split case, let $g_{\bm{x}_{[i-1]}}=k$ and $\rho_{i}^*=\rho_i\cup S_{M_i+1}$, where $g_{\bm{x}_{[i-1]}'}=M_i+1$ and, for any other $\tilde{\bm{x}}_{[i-1]}\in S_k$, $P(g_{\tilde{\bm{x}}_{[i-1]}}=k)=P(g_{\tilde{\bm{x}}_{[i-1]}}=M_i+1)=0.5$. In other words, a merge move combines the two selected stages into a single block, while a split move creates a new stage by relocating one of the selected contexts and randomly assigning the remaining members of the original stage to one of the two resulting blocks. The move is then either accepted or rejected using a Metropolis-Hastings step. The probability of accepting the move equals 
\begin{equation}
\label{eq:metropolis}
\min\left\{1,\exp\left(\log \frac{P(\mathcal{D}|\mathcal{T}_{\bm{X}}^{\rho_i^*})}{P(\mathcal{D}|\mathcal{T}_{\bm{X}}^{\rho_i})} + \log \frac{P(\mathcal{T}_{\bm{X}}^{\rho_i^*})}{P(\mathcal{T}_{\bm{X}}^{\rho_i})} + \log \frac{P(\mathcal{T}_{\bm{X}}^{\rho_i}|\mathcal{T}_{\bm{X}}^{\rho_i^*})}{P(\mathcal{T}_{\bm{X}}^{\rho_i^*}|\mathcal{T}_{\bm{X}}^{\rho_i})}\right)\right\}
\end{equation}
The three ratios in Equation \eqref{eq:metropolis} can be easily computed noticing that the two trees are \textit{nested} \citep{collazo2016new}: one can be obtained from the other by either merging two of its stages or splitting one into two. Using this fact, it follows that, in the case of a merge move
\begin{align}
\log \frac{P(\mathcal{D}|\mathcal{T}_{\bm{X}}^{\rho_i^*})}{P(\mathcal{D}|\mathcal{T}_{\bm{X}}^{\rho_i})} &= \log m(\bm{N}_{S_k\cup S_{k'}}) - \log m(\bm{N}_{S_k}) - \log m(\bm{N}_{S_{k'}}),\label{eq:rati}\\
\log \frac{P(\mathcal{T}_{\bm{X}}^{\rho_i^*})}{P(\mathcal{T}_{\bm{X}}^{\rho_i})} & = \log\kappa + \log\frac{\Gamma(\#\{S_k\cup S_{k'}\})}{\Gamma(\#S_k)\Gamma(\#S_{k'})}-\xi\sum_{j\in S_k,l\in S_{k'}}d_{j,l} -\sum\limits_{z=1}^q \lambda_z \sum_{j\in S_k,l\in S_{k'}} \delta^{(z)}_{j,l},\nonumber \\
\log \frac{P(\mathcal{T}_{\bm{X}}^{\rho_i}|\mathcal{T}_{\bm{X}}^{\rho_i^*})}{P(\mathcal{T}_{\bm{X}}^{\rho_i^*}|\mathcal{T}_{\bm{X}}^{\rho_i})} &= \log0.5 \cdot (\# \{S_k\cup S_{k'}\}-2).\nonumber
\end{align}
Equation \eqref{eq:rati} can be further simplified to
\begin{equation}
\label{eq:ciao}
    \frac{P(\mathcal{D}|\mathcal{T}_{\bm{X}}^{\rho_i^*})}{P(\mathcal{D}|\mathcal{T}_{\bm{X}}^{\rho_i})} = \frac{|\bm{N}_{S_k}|!|\bm{N}_{S_{k'}}|!}{|\bm{N}_{S_k\cup S_{k'}}|!}\cdot \prod_{x_i\in\mathbb{X}_i}\frac{N_{S_k\cup S_{k'}}^{x_i}!}{N_{S_k}^{x_i}!N_{S_{k'}}^{x_i}!}
\end{equation}
The proof of this equation is in Supplementary Section~\ref{SMsec:proof2}. Analogous expressions can be derived for the split move by symmetry.

\subsection{Proofs}
\subsubsection{Proof of Equation~(\ref{eq:fact})}\label{SMsec:proof1}

Recall that the function $m(\bm{N}_S)$ corresponding to the marginal likelihood contribution from each stage and takes the form:
\begin{equation}
\label{eq:marginal2}
    \log m(\bm{N}_S) = 
    \log \Gamma(|\bm{a}_S|) 
    - \log \Gamma(|\bm{a}_S + \bm{N}_S|) 
    + |\log \Gamma(\bm{a}_S + \bm{N}_S)| 
    - |\log \Gamma(\bm{a}_S)|,
\end{equation}
In our setup both $S_{k}\cup\bm{x}_{[i-1]}$ and $S_{k}\setminus \bm{x}_{[i-1]}$ are given the prior $a/\#\mathbb{X}_i$. Therefore all terms in Equation (\ref{eq:marginal2}) involving only the hyperparameter $\bm{a}$ cancel out. Hence
\[
\frac{m\left(\bm{N}_{S_k\cup\bm{x}_{[i-1]}}\right)}{m\left(\bm{N}_{S_k\setminus \bm{x}_{[i-1]}}\right)}=\frac{\prod_{x_i\in\mathbb{X}_i}\Gamma(a/\#\mathbb{X}_i+N_{S_k\cup \bm{x}_{[i-1]}}^{x_i})}{\prod_{x_i\in\mathbb{X}_i}\Gamma(a/\#\mathbb{X}_i+N_{S_k\setminus \bm{x}_{[i-1]}}^{x_i})}\cdot\frac{\Gamma(a+|\bm{N}_{S_k\setminus\bm{x}_{[i-1]}}|)}{\Gamma(a+|\bm{N}_{S_k\cup\bm{x}_{[i-1]}}|)}
\]
Noticing that $N_{S_k\cup\bm{x}_{[i-1]}}^{x_i}=N_{S_k\setminus\bm{x}_{[i-1]}}^{x_i}+N_{\bm{x}_{[i-1]}}^{x_i}$ and recalling that $\Gamma(t+1)=t\Gamma(t)$ for any $t>0$, we have that
\begin{align*}
\frac{\prod_{x_i\in\mathbb{X}_i}\Gamma(a/\#\mathbb{X}_i+N_{S_k\cup \bm{x}_{[i-1]}}^{x_i})}{\prod_{x_i\in\mathbb{X}_i}\Gamma(a/\#\mathbb{X}_i+N_{S_k\setminus \bm{x}_{[i-1]}}^{x_i})}&=\prod_{x_i\in\mathbb{X}_i}N_{\bm{x}_{[i-1]}}^{x_i}!
\\
\frac{\Gamma(a+|\bm{N}_{S_k\setminus\bm{x}_{[i-1]}}|)}{\Gamma(a+|\bm{N}_{S_k\cup\bm{x}_{[i-1]}}|)}&=\frac{1}{|\bm{N}_{\bm{x}_{[i-1]}}|!}
\end{align*}
and the result follows.

\subsubsection{Proof of Equation~(\ref{eq:ciao})}\label{SMsec:proof2}

From Equation \eqref{eq:marginal1} we have that, using the fact that the hyperparameters $\bm{a}=a/\#\mathbb{X}_i$
\begin{align*}
    m(\bm{N}_{S_k})&=\frac{\Gamma(a)}{\Gamma(a/\#\mathbb{X}_i)^{\#\mathbb{X}_i}}\frac{\prod_{x_i\in\mathbb{X}_i}\Gamma(a/\#\mathbb{X}_i+N_{S_{k}}^{x_i})}{\Gamma(a+|\bm{N}_{S_k}|)}\\
    m(\bm{N}_{S_{k'}})&=\frac{\Gamma(a)}{\Gamma(a/\#\mathbb{X}_i)^{\#\mathbb{X}_i}}\frac{\prod_{x_i\in\mathbb{X}_i}\Gamma(a/\#\mathbb{X}_i+N_{S_{k'}}^{x_i})}{\Gamma(a+|\bm{N}_{S_{k'}}|)}\\
     m(\bm{N}_{S_{k'}\cup S_k})&=\frac{\Gamma(a)}{\Gamma(a/\#\mathbb{X}_i)^{\#\mathbb{X}_i}}\frac{\prod_{x_i\in\mathbb{X}_i}\Gamma(a/\#\mathbb{X}_i+N_{S_{k'}\cup S_k}^{x_i})}{\Gamma(a+|\bm{N}_{S_{k'}\cup S_k}|)}
\end{align*}
By recursively using $\Gamma(t+1)=t\Gamma(t)$, we for instance have that
\[
\Gamma(a+|\bm{N}_{S_k}|)=\Gamma(a)|\bm{N}_{S_k}|! \,\,\,\mbox{ and } \,\,\,\Gamma(a/\#\mathbb{X}_i+N_{S_{k}}^{x_i}) = \Gamma(a/\#\mathbb{X}_i)N_{S_k}^{x_i}!
\]
Hence
\begin{align*}
\frac{m(\bm{N}_{S_k\cup S_{k'}})}{m(\bm{N}_{S_k})m(\bm{N}_{S_{k'}})}&=\frac{\Gamma(a/\#\mathbb{X}_i)^{\#\mathbb{X}_i}}{\Gamma(a)}\cdot\frac{\prod_{x_i\in\mathbb{X}_i}N_{S_k\cup S_{k'}}^{x_i}!}{\prod_{x_i\in\mathbb{X}_i}\Gamma(a/\#\mathbb{X}_i)N_{S_k}^{x_i}!N_{S_{k'}}^{x_i}!}\cdot\frac{\Gamma(a)|\bm{N}_{S_k}|!|\bm{N}_{S_k'}|!}{|\bm{N}_{S_k\cup S_{k'}}|!}\\
&=\frac{|\bm{N}_{S_k}|!|\bm{N}_{S_{k'}}|!}{|\bm{N}_{S_k\cup S_{k'}}|!}\cdot \prod_{x_i\in\mathbb{X}_i}\frac{N_{S_k\cup S_{k'}}^{x_i}!}{N_{S_k}^{x_i}!N_{S_{k'}}^{x_i}!}
\end{align*}

\clearpage
\subsection{Additional Figures and Tables}\label{SMsec:Additional_FigsTabs}

\begin{table}[ht]
    \caption{Prior distribution for the tree in Figure 1 
    over the vertices: $\bm{x}_{(0,0)}$ ($a$), $\bm{x}_{(0,1)}$ ($b$), $\bm{x}_{(1,0)}$ ($c$), and $\bm{x}_{(1,1)}$ ($d$), for $\kappa=0.5,1$ and $\xi=0.2,0.8$.}
    \label{SMtable:pars}
    \centering
    \scriptsize
    \begin{tabular}{cccc}
    \toprule
    \textbf{Partition} & \textbf{\(\kappa\)} & \textbf{Prob (\(\xi = 0.2\))} & \textbf{Prob (\(\xi = 0.8\))} \\
    \midrule
    \multirow{2}{*}{$\{a\},\{b\},\{c\},\{d\}$} 
        & \(\kappa = 0.5\) & 0.024 & 0.117\\
        & \(\kappa = 1\)   & 0.082&0.251 \\
    \multirow{2}{*}{$\{a,b\},\{c\},\{d\}$}     
        & \(\kappa = 0.5\) & 0.039 & 0.105 \\
        & \(\kappa = 1\)   &0.067 & 0.113\\
    \multirow{2}{*}{$\{a,c\},\{b\},\{d\}$}     
        & \(\kappa = 0.5\) & 0.039 & 0.105 \\
        & \(\kappa = 1\)   & 0.067& 0.113\\
    \multirow{2}{*}{$\{a,d\},\{b\},\{c\}$}     
        & \(\kappa = 0.5\) & 0.032 & 0.047 \\
        & \(\kappa = 1\)   &0.055 & 0.051\\
    \multirow{2}{*}{$\{a\},\{b,c\},\{d\}$}     
        & \(\kappa = 0.5\) &0.032 & 0.047 \\
        & \(\kappa = 1\)   &0.055 &0.051 \\
    \multirow{2}{*}{$\{a\},\{b,d\},\{c\}$}     
        & \(\kappa = 0.5\) & 0.039 & 0.105 \\
        & \(\kappa = 1\)   & 0.067 & 0.113 \\
    \multirow{2}{*}{$\{a\},\{b\},\{c,d\}$}     
        & \(\kappa = 0.5\) & 0.039 & 0.105 \\
        & \(\kappa = 1\)   & 0.067 & 0.113 \\
    \multirow{2}{*}{$\{a,b\},\{c,d\}$}         
        & \(\kappa = 0.5\) &  0.065& 0.094\\
        & \(\kappa = 1\)   &0.055 & 0.051 \\
    \multirow{2}{*}{$\{a,c\},\{b,d\}$}         
        & \(\kappa = 0.5\) & 0.065& 0.094 \\
        & \(\kappa = 1\)   &0.055 & 0.051 \\
    \multirow{2}{*}{$\{a,d\},\{b,c\}$}         
        & \(\kappa = 0.5\) & 0.043& 0.019 \\
        & \(\kappa = 1\)   & 0.037& 0.010\\
    \multirow{2}{*}{$\{a,b,c\},\{d\}$}         
        & \(\kappa = 0.5\) & 0.087 & 0.038 \\
        & \(\kappa = 1\)   & 0.074 &0.020 \\
    \multirow{2}{*}{$\{a,b,d\},\{c\}$}         
        & \(\kappa = 0.5\) & 0.087& 0.038 \\
        & \(\kappa = 1\)   &0.074 &0.020 \\
    \multirow{2}{*}{$\{a,c,d\},\{b\}$}         
        & \(\kappa = 0.5\) &0.087& 0.038 \\
        & \(\kappa = 1\)   & 0.074 &0.020 \\
    \multirow{2}{*}{$\{a\},\{b,c,d\}$}         
        & \(\kappa = 0.5\) &0.087& 0.038\\
        & \(\kappa = 1\)   & 0.074 &0.020 \\
    \multirow{2}{*}{$\{a,b,c,d\}$}             
        & \(\kappa = 0.5\) & 0.234 & 0.009 \\
        & \(\kappa = 1\)   & 0.099 & 0.003 \\
    \bottomrule
    \end{tabular}
\end{table}

\begin{figure}
 \centering
    \includestandalone[width=0.7\linewidth]{stage_simul}
    \caption{Simulated data. The Figure shows the staged tree employed to simulate data in Section 3. 
    Each edge is labeled with the corresponding transition probability. For simplicity, colors are reused across different depths, but stages should be interpreted within each depth independently.}
    \label{SMfig:stage_simul}
\end{figure}

\begin{figure}
    \centering
    \includegraphics[width=0.85\linewidth]{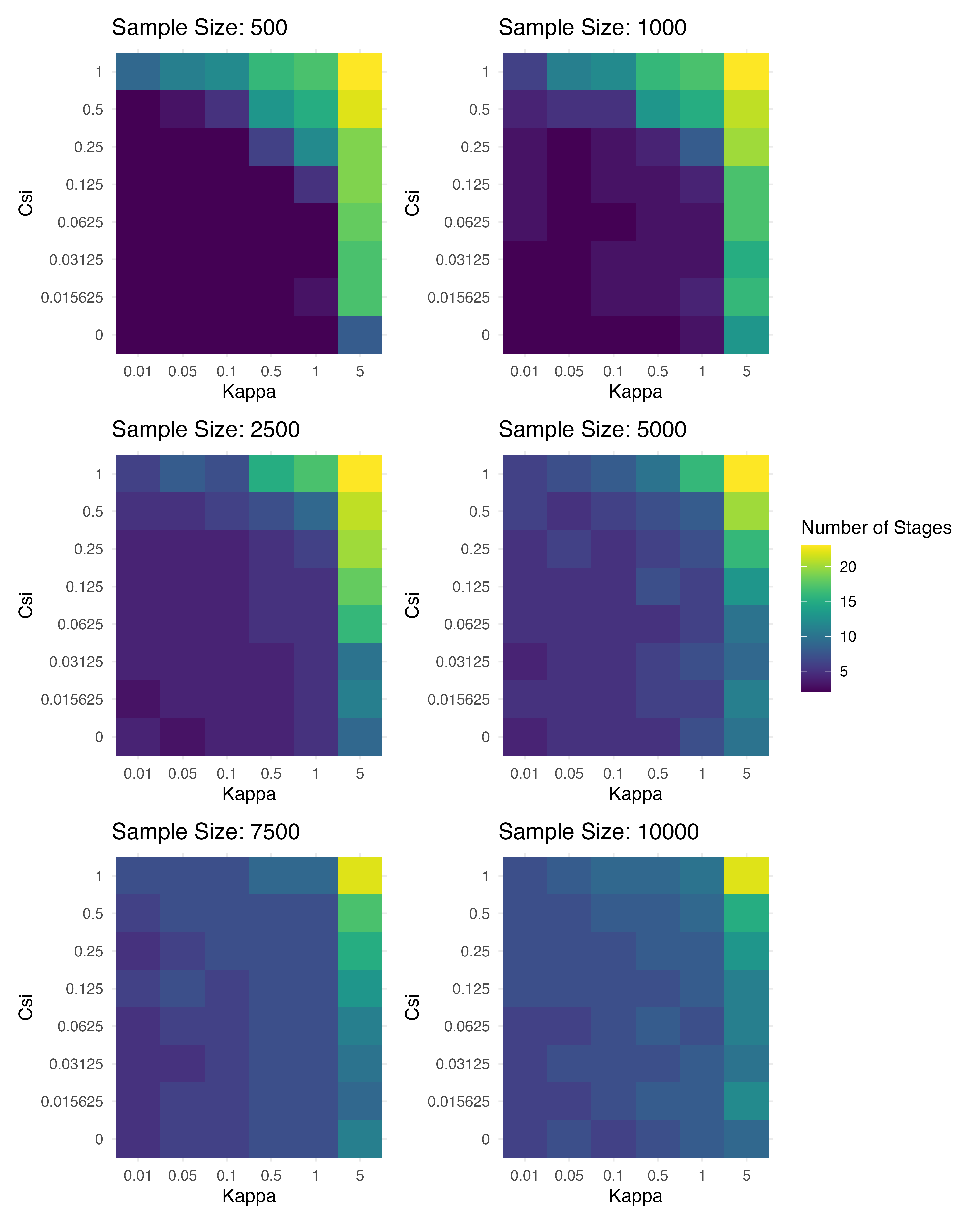}
\caption{Simulated data. Median number of stages estimated for the last variable in the staged tree, across combinations of prior hyperparameters $\kappa$ and $\xi$, and increasing sample sizes ($n$). Results are averaged over 5 replications per setting.}
    \label{SMfig:sta}
\end{figure}

\begin{figure}
    \centering
    \begin{subfigure}{0.99\linewidth}
        \centering
        \includegraphics[width=\linewidth]{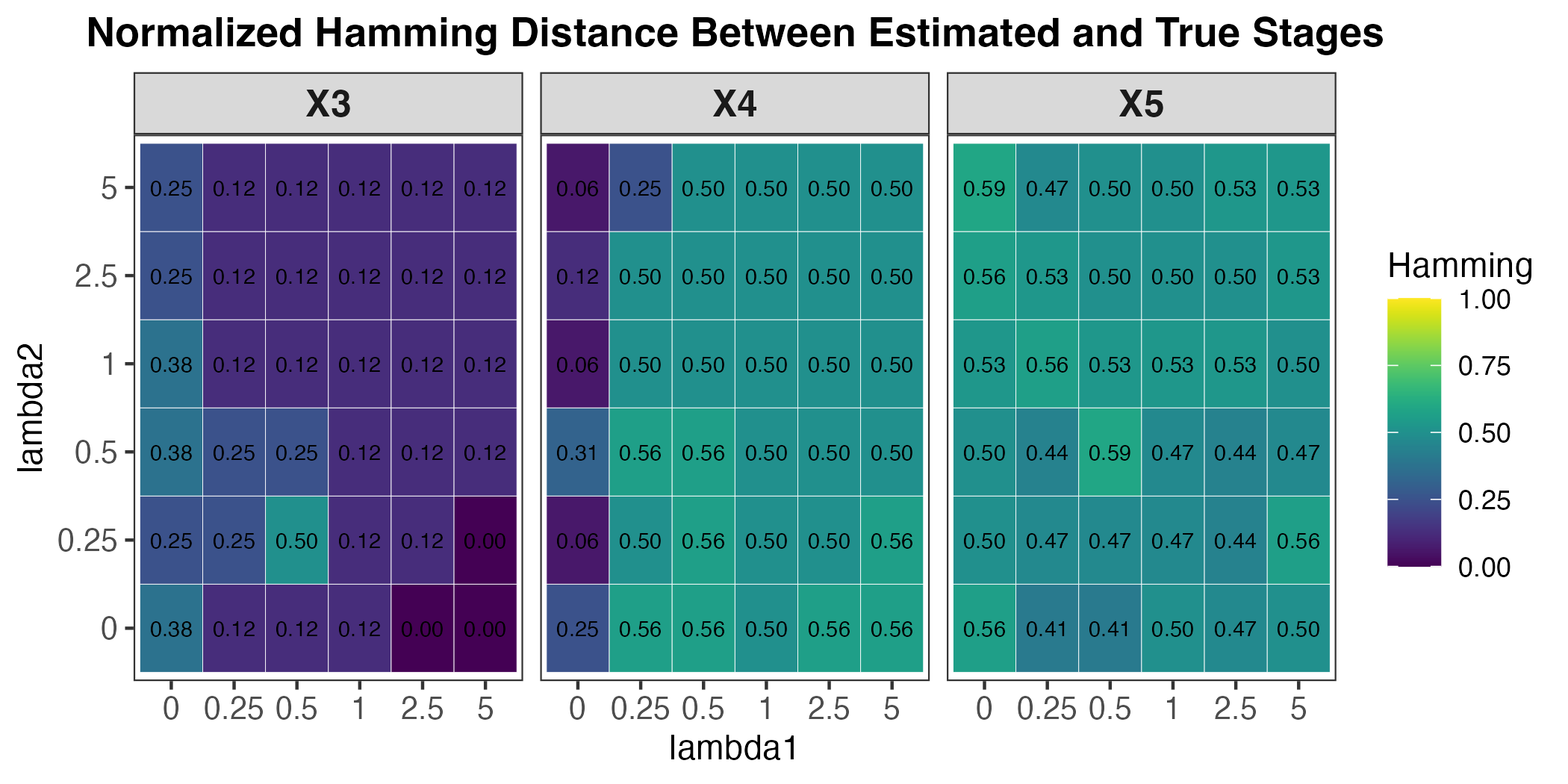}
        \caption{Hamming distance between estimated and true stage assignments.}
        \label{SMfig:hamming_heatmap}
    \end{subfigure}
    
    \begin{subfigure}{0.99\linewidth}
        \centering
        \includegraphics[width=\linewidth]{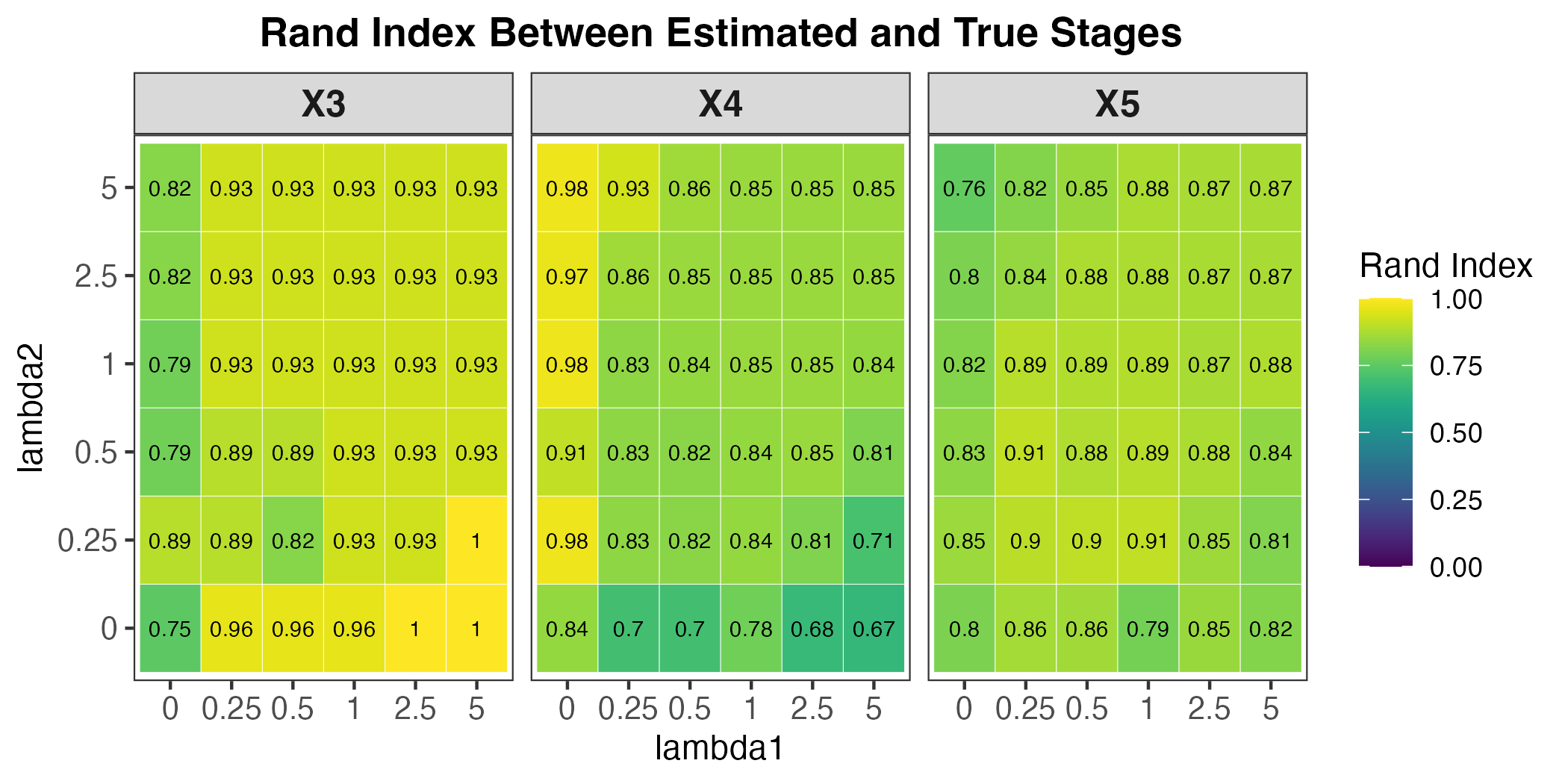}
        \caption{Rand index between estimated and true stage assignments.}
        \label{SMfig:rand_heatmap}
    \end{subfigure}
    
    \caption{Simulated data. Evaluation of stage recovery across combinations of $(\lambda_1, \lambda_2)$.}
    \label{SMfig:heatmap_comparison}
\end{figure}

\begin{figure}[htbp]
    \centering
    \includegraphics[width=0.7\linewidth]{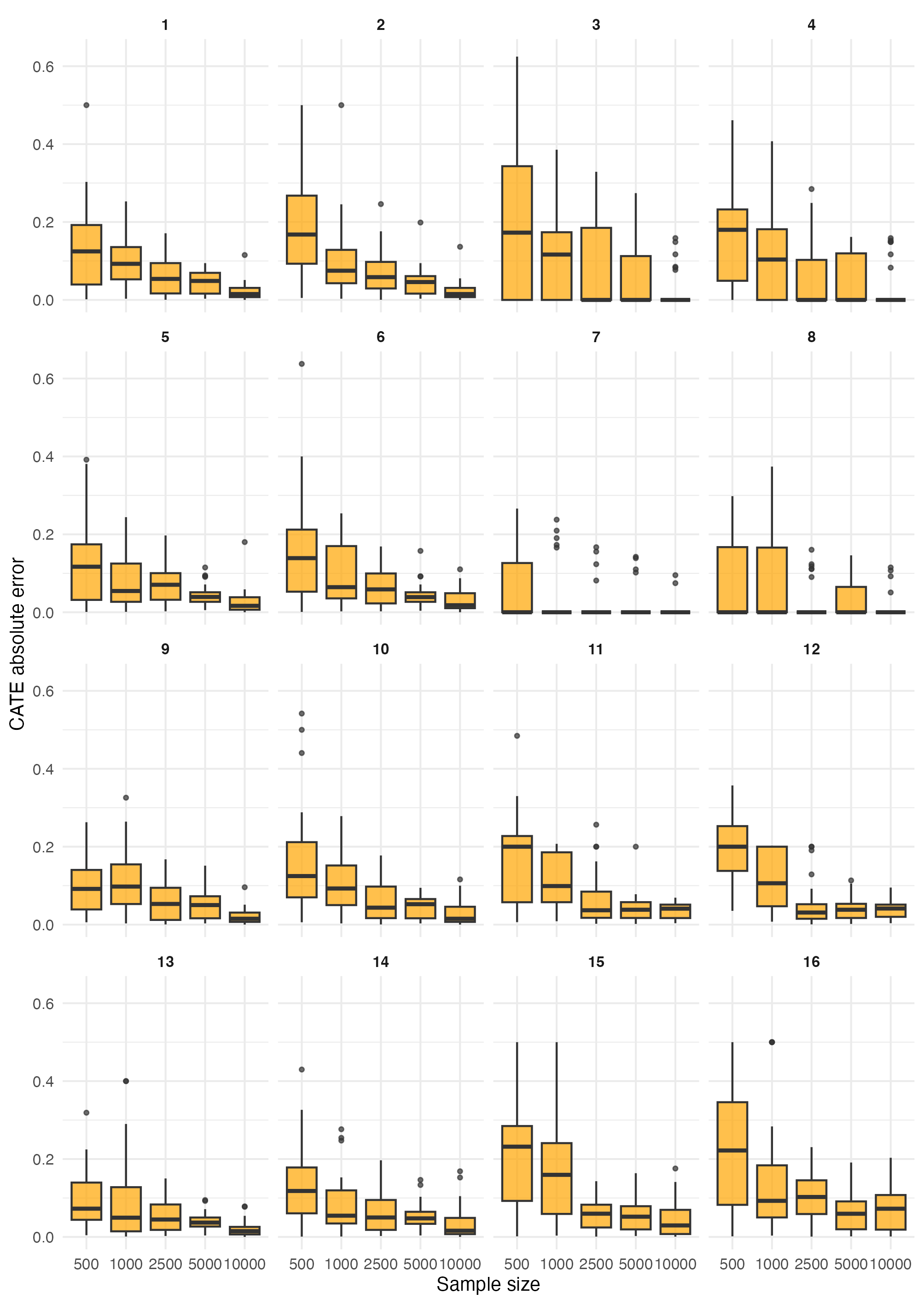}
    \caption{Simulated data. Absolute estimation error of the CATEs across the 16 covariate profiles.}
    \label{fig:cate_sim}
\end{figure}

\begin{table}[ht]
    \centering
    \begin{tabular}{c c c c c c c c c}
        \toprule
        & V1 & V2 & V3 & V4 & V5 & V6 & V7 & V8 \\
        \midrule
        \multicolumn{9}{c}{\emph{Vertical Lower Bound}} \\
        L1 & 1 & 2 & 3 & 1 & 4 & 5 & 6 & 7 \\
        L2 & 1 & 2 & 3 & 4 & 4 & 5 & 6 & 7 \\
        L3 & 1 & 2 & 3 & 4 & 5 & 6 & 7 & 4 \\
        \midrule
        \multicolumn{9}{c}{\emph{Vertical Upper Bound}} \\
        U1 & 1 & 2 & 1 & 1 & 3 & 4 & 1 & 1 \\
        \midrule
        \multicolumn{9}{c}{\emph{Horizontal Bound}} \\
        H1  & 1 & 2 & 1 & 2 & 3 & 4 & 5 & 2 \\
        H2  & 1 & 2 & 3 & 2 & 4 & 5 & 3 & 2 \\
        H3  & 1 & 2 & 3 & 2 & 4 & 5 & 1 & 1 \\
        H4  & 1 & 2 & 1 & 2 & 3 & 4 & 5 & 2 \\
        H5  & 1 & 2 & 1 & 2 & 3 & 4 & 5 & 2 \\
        H6  & 1 & 2 & 3 & 2 & 4 & 5 & 3 & 2 \\
        H7  & 1 & 2 & 1 & 2 & 3 & 4 & 5 & 2 \\
        H8  & 1 & 2 & 1 & 2 & 3 & 4 & 5 & 2 \\
        H9  & 1 & 2 & 3 & 2 & 4 & 5 & 3 & 2 \\
        H10 & 1 & 2 & 3 & 2 & 4 & 5 & 3 & 2 \\
        H11 & 1 & 2 & 1 & 2 & 3 & 4 & 5 & 2 \\
        H12 & 1 & 2 & 3 & 2 & 4 & 5 & 1 & 2 \\
        H13 & 1 & 2 & 1 & 2 & 3 & 4 & 5 & 2 \\
        H14 & 1 & 2 & 1 & 2 & 3 & 4 & 5 & 1 \\
        H15 & 1 & 2 & 1 & 1 & 3 & 4 & 5 & 2 \\
        H16 & 1 & 2 & 3 & 2 & 4 & 5 & 1 & 1 \\
        \bottomrule
    \end{tabular}
        \caption{Cesarean Section data. Credible ball partitions for \texttt{monitor}. Each row corresponds to a different staging: the vertical lower bound (top 3 rows), vertical upper bound (next row), and horizontal bound (last 16 rows).}
    \label{SMtab:monitor_ball}
\end{table}

\begin{table}[ht]
    \centering
    \scalebox{0.85}{
    \begin{tabular}{c c c c c c c c c c c c c c c c c}
        \toprule
        & V1 & V2 & V3 & V4 & V5 & V6 & V7 & V8 & V9 & V10 & V11 & V12 & V13 & V14 & V15 & V16 \\
        \midrule
        \multicolumn{17}{c}{\emph{Vertical Lower Bound}} \\
        L1 & 1 & 2 & 3 & 4 & 5 & 3 & 6 & 7 & 8 & 8 & 9 & 6 & 6 & 4 & 7 & 10 \\
        L2 & 1 & 2 & 3 & 4 & 3 & 3 & 5 & 5 & 6 & 6 & 7 & 8 & 5 & 7 & 9 & 5 \\
        L3 & 1 & 2 & 3 & 4 & 3 & 3 & 4 & 5 & 6 & 6 & 4 & 7 & 8 & 4 & 9 & 5 \\
        L4 & 1 & 2 & 3 & 4 & 5 & 3 & 4 & 6 & 7 & 7 & 4 & 8 & 9 & 5 & 10 & 6 \\
        \midrule
        \multicolumn{17}{c}{\emph{Vertical Upper Bound}} \\
        U1 & 1 & 2 & 3 & 4 & 3 & 3 & 4 & 5 & 6 & 6 & 4 & 4 & 4 & 4 & 5 & 5 \\
        \midrule
        \multicolumn{17}{c}{\emph{Horizontal Bound}} \\
        H1 & 1 & 2 & 3 & 4 & 5 & 3 & 6 & 7 & 8 & 8 & 9 & 6 & 6 & 4 & 7 & 10 \\
        H2 & 1 & 2 & 3 & 4 & 3 & 3 & 5 & 5 & 6 & 6 & 7 & 8 & 5 & 7 & 9 & 5 \\
        H3 & 1 & 2 & 3 & 4 & 3 & 3 & 4 & 5 & 6 & 6 & 4 & 7 & 8 & 4 & 9 & 5 \\
        H4 & 1 & 2 & 3 & 4 & 5 & 3 & 4 & 6 & 7 & 7 & 4 & 8 & 9 & 5 & 10 & 6 \\
        \bottomrule
    \end{tabular}}
        \caption{Cesarean Section data. Credible ball partitions for \texttt{cesarean}. Each row corresponds to a different staging: the vertical lower bound (top 4 rows), vertical upper bound (1 row), and horizontal bound (4 rows).}
    \label{SMtab:cesarean_ball}
\end{table}

\end{document}